\pdfoutput=1
\documentclass[11pt]{article}

\newcommand{\newc}{\newcommand}
\def\eq$#1${\begin{equation}#1\end{equation}}
\def\gat$#1${\begin{gather}#1\end{gather}}
\def\bal$#1${\begin{align}#1\end{align}}
\def\eqarr$#1${\begin{eqnarray}#1\end{eqnarray}}
\newc{\pa}{\partial}
\newc{\alp}{\alpha}
\newc{\gam}{\gamma}
\newc{\Gam}{\Gamma}
\newc{\del}{\delta}
\newc{\eps}{\epsilon}
\newc{\lam}{\lambda}
\newc{\sig}{\sigma}
\newc{\ups}{\upsilon}
\newc{\ome}{\omega}
\newc{\nonum}{\nonumber}
\newc{\vph}{\varphi}
\newc{\tx}{\tilde{x}}
\newc{\paper}[2]{\enquote{\textit{#1}} \cite{#2}}
\newc{\erf}{\text{erf}}
\newcommand{\myref}[2]{{\color{MyDarkBlue}\ref{#1}(\subref{#2})}} 

\usepackage{graphicx,rotating,colortbl}
\usepackage{jcappub}
\usepackage{amsmath}
\usepackage{amssymb}
\usepackage{braket}
\usepackage{graphicx}
\usepackage{hyperref}
\usepackage{mathrsfs}
\usepackage{subcaption}
\usepackage{empheq}
\usepackage{appendix}
\usepackage{natbib}
\usepackage[dvipsnames]{xcolor}
\usepackage{float}\usepackage[top=1in, bottom=0.4in, left=0.8in, right=0.8in, includefoot]{geometry} 
\usepackage[autostyle]{csquotes}
\usepackage{color,soul}
\usepackage{mathtools}
\usepackage{multirow}
\usepackage{tabu}
\usepackage{fixmath}

\usepackage{ulem} 

\usepackage{titlesec}

\titlespacing\section{0pt}{12pt plus 4pt minus 2pt}{5pt plus 2pt minus 2pt}
\titlespacing\subsection{0pt}{12pt plus 4pt minus 2pt}{12pt plus 2pt minus 2pt}

\definecolor{azure(colorwheel)}{rgb}{0.0, 0.5, 1.0}
\definecolor{DarkViolet}{RGB}{148,0,211}
\definecolor{MyDarkBlue}{rgb}{0,0.1,0.7}
\definecolor{DarkBlue}{RGB}{0,0,153}
\definecolor{amber}{rgb}{1.0, 0.49, 0.0}
\definecolor{amaranth}{rgb}{0.9, 0.17, 0.31}
\definecolor{nicered}{rgb}{0.7,0.1,0.1}
\definecolor{brown}{rgb}{0.5,0.1,0.1}
\definecolor{nicegreen}{rgb}{0.0,0.3,0.0}
\definecolor{tealgreen}{rgb}{0.0, 0.51, 0.5}
\hypersetup{colorlinks,bookmarksopen,
bookmarksnumbered,
citecolor={nicered},
linkcolor={MyDarkBlue},
urlcolor={tealgreen},
pdfstartview=FitH}

\usepackage{courier}

\graphicspath{{Figs/}}
\newcommand{\comm}[1]{}

\newc{\teo}[1]{\textcolor{azure(colorwheel)}{#1}} 
\newc{\com}[1]{\textcolor{amaranth}{#1}} 
\newc{\bako}[1]{\textcolor{DarkViolet}{#1}} 
\newc{\corr}[1]{\textcolor{red}{#1}} 

\usepackage{scalerel}
\usepackage{tikz}
\usetikzlibrary{svg.path}
\definecolor{orcidlogocol}{HTML}{A6CE39}
\tikzset{
  orcidlogo/.pic={
    \fill[orcidlogocol] svg{M256,128c0,70.7-57.3,128-128,128C57.3,256,0,198.7,0,128C0,57.3,57.3,0,128,0C198.7,0,256,57.3,256,128z};
    \fill[white] svg{M86.3,186.2H70.9V79.1h15.4v48.4V186.2z}
                 svg{M108.9,79.1h41.6c39.6,0,57,28.3,57,53.6c0,27.5-21.5,53.6-56.8,53.6h-41.8V79.1z M124.3,172.4h24.5c34.9,0,42.9-26.5,42.9-39.7c0-21.5-13.7-39.7-43.7-39.7h-23.7V172.4z}
                 svg{M88.7,56.8c0,5.5-4.5,10.1-10.1,10.1c-5.6,0-10.1-4.6-10.1-10.1c0-5.6,4.5-10.1,10.1-10.1C84.2,46.7,88.7,51.3,88.7,56.8z};}}
\newcommand\orcid[1]{\href{https://orcid.org/#1}{\mbox{\scalerel*{
\begin{tikzpicture}[yscale=-1,transform shape]
\pic{orcidlogo};
\end{tikzpicture}
}{|}}}}

\title{Analytic and asymptotically flat hairy (ultra-compact) black-hole solutions and their \\axial perturbations}

\author[a,b]{Athanasios Bakopoulos\orcid{0000-0002-3012-6144}}
\author{and}
\author[a]{Theodoros Nakas\orcid{0000-0002-3522-5803}}

\emailAdd{a.bakop@uoi.gr}
\emailAdd{t.nakas@uoi.gr}

\affiliation[a]{Division of Theoretical Physics, Department of Physics, University of Ioannina, 
Ioannina GR-$45110$, Greece\\}
\affiliation[b]{Division of Applied Analysis, Department of Mathematics,
University of Patras, Rio Patras GR-26504, Greece\\}

\abstract{In this work, we consider a very simple gravitational theory that contains a scalar field with its kinetic and potential terms minimally coupled to gravity, while the scalar field is assumed to have a coulombic form. In the context of this theory, we study an analytic, asymptotically flat, and regular (ultra-compact) black-hole solutions with non-trivial scalar hair of secondary type. At first, we examine the properties of the static and spherically symmetric black-hole solution---firstly appeared in \href{https://arxiv.org/abs/1504.08209}{1504.08209 [gr-qc]}---and we find that in the causal region of the spacetime the stress-energy tensor, needed to support our solution, satisfies the strong energy conditions. Then, by using the slow-rotating approximation, we generalize the static solution into a slowly rotating one, and we determine explicitly its angular velocity $\omega(r)$. We also find that the angular velocity of our ultra-compact solution is always larger compared to the angular velocity of the corresponding equally massive slow-rotating Schwarzschild black hole. In addition, we investigate the axial perturbations of the derived solutions by determining the Schr\"{o}dinger-like equation and the effective potential. We show that there is a region in the parameter space of the free parameters of our theory, which allows for the existence of stable ultra-compact black hole solutions. Specifically, we calculate that the most compact and stable black hole solution is 0.551 times smaller than the Schwarzschild one, while it rotates 2.491 times faster compared to the slow-rotating Schwarzschild black hole. Finally, we present without going into details the generalization of the derived asymptotically flat solutions to asymptotically (A)dS solutions.}

\begin{document}

\maketitle
\vspace*{0.8cm}

\section{Prologue}
\label{intro}

General theory of Relativity (GR) is now more than a hundred years old, and during this time interval, it has been tested countless times at different distance scales. It counts many experimental successes with the most recent ones being the detection of the gravitational waves \cite{Abbott:2016blz} and the black hole shadow \cite{Akiyama:2019cqa}. However, cosmological observations indicate that General Relativity is an effective theory and is therefore expected to break down at some distance or energy scale. Indeed, the Standard Model for Cosmology has many open problems, like the nature of dark energy and dark matter or the accurate model for inflation. Therefore, these observations motivate the need for modified gravitational theories. The simplest and most studied modified gravitational theories are the Scalar Tensor (ST) theories where the scalar field provides an additional scalar degree of freedom. One important property of scalar tensor theories is that most modified gravitational theories have a limit in which they reduce to scalar-tensor theories. Moreover, after the detection of the Higgs boson in 2012 \cite{Aad:2012tfa,Chatrchyan:2012ufa} it is now widely accepted that scalar fields exist in nature. Consequently, scalar-tensor theories constitute a very fertile framework in which new ideas can be easily applied and new spacetime geometries can be easily generated. 

Although the Modified Gravitational theories are usually constructed as cosmological models, the existence of local (black-hole, neutron star or even star) solutions  is  essential  for their credibility. If it is not possible to give rise to astrophysically realistic local solutions, then the theory cannot be considered viable. However, the search for new black hole solutions in the scalar tensor theories was  prematurely stopped due to the formulation of No-Scalar Hair theorems \cite{NH-scalar1, NH-scalar2} that forbade the existence of a static black hole solution associated with a non-trivial scalar field.  Nevertheless, the validity of the first no-scalar hair theorem was disputed after the derivation of new black-hole solutions. Some indicative examples are solutions  with  Yang-Mills gauge fields \cite{YM1,YM2,YM3,YM4},  Skyrme fields \cite{Skyrmions1,Skyrmions2} or with a conformal coupling to gravity \cite{Conformal1, Bekenstein:1975ts}.  Bekenstein developed a new formulation of the no scalar hair theorem at 1995 \cite{Bekenstein} but this was also shown to be evaded only after a year with the derivation of the dilatonic Gauss-Bonnet black holes \cite{Kanti:1995vq}. In addition, the revival of the Horndeski theory \cite{Horndeski} through the generalized Galileon theory \cite{Galileon} has significantly enhanced the number of new black hole solutions in scalar tensor theories. The last decades a vast number of hairy black-hole solutions has appeared in the literature; for a partial list of asymptotically flat solutions see Refs. \cite{Torii:1993vm, Bechmann:1995sa, Dennhardt:1996cz, Nucamendi:1995ex, Gubser:2005ih, Bronnikov:2005gm, Nikonov_2008, Anabalon:2012ih, Anabalon:2013qua, Kleihaus:2013tba, Babichev:2013cya, Sotiriou:2014pfa, Herdeiro:2014goa, Charmousis:2014zaa, Astorino:2014mda, Cadoni:2015gfa, Herdeiro:2015gia, Kleihaus:2015iea, Tahamtan:2015sra, Tolley:2015ywa, Hod:2017kpt, Herdeiro:2016tmi, Ni:2016rhz, Benkel:2016rlz, Sanchis-Gual:2016tcm, Heisenberg:2017xda, Antoniou:2017acq, Antoniou:2017hxj, Herdeiro:2018wvd, Pacilio:2018gom, Brihaye:2018woc, Astefanesei:2019mds, Brihaye:2018grv, Wang:2018xhw, Herdeiro:2018daq, VanAelst:2019kku, Hod:2019pmb, Kunz:2019sgn, Cunha:2019dwb, Filippini:2019cqk, Zou:2019ays, Fernandes:2020gay, Santos:2020pmh, Sultana:2020pcc, Hong:2020miv, Herdeiro:2020wei, Astefanesei:2020xvn, Shnir:2020hau, Hod:2020jjy, Ovalle:2020kpd, Brihaye:2021ich, Delgado:2020hwr, Myung:2020ctt,Doneva:2017bvd,Silva:2017uqg}, while for asymptotically (A)dS$_4$ solutions see Refs. \cite{Martinez:2004nb, Martinez:2006an, Anabalon:2012ta, Charmousis:2015aya, Babichev:2015rva, Fan:2015oca, Perapechka:2016cof, Bakopoulos:2018nui, BenAchour:2018dap, Brihaye:2019gla, Guo:2021zed}. Also, in the recent review \cite{Faraoni:2021nhi}, one can find inhomogeneous solutions in the context of both GR and scalar-tensor gravity (see also \cite{Bakopoulos:2022csr}). It is important to mention though, that the existence of black hole solutions does not guarantee by itself that these black holes are serious candidates for astrophysical objects.  An astrophysical black hole must be stable and also it must rotate.  Hence, it is crucial to study the existence  of local solutions in scalar-tensor theories along with their rotation and stability. 

Furthermore, in scalar-tensor theories we may derive solutions for local objects that are forbidden in General Relativity like wormholes or particle-like solutions (solitons). In General Relativity, wormholes require the existence of exotic matter near their throat \cite{Morris:1988cz}, that violates the energy conditions, while in electrovacuum the solitons are unstable \cite{Wheeler:1955zz, Misner:1957mt}. In scalar-tensor theories though, the scalar field generates an effective energy momentum tensor that may violate the energy conditions and generate wormhole solutions, while in the same time the ordinary matter that may exist around the throat remains non-exotic.  Although this is mainly achieved using phantom scalar fields, i.e. scalar fields that have a negative sign in front of their kinetic term, it has been shown that in the Einstein-scalar-Gauss-Bonnet theory  real scalar fields may support regular wormhole solutions \cite{Kanti:2011jz,Kanti:2011yv,Antoniou:2019awm}. Regular wormhole solutions may be also found in the framework of the beyond Horndeski theories \cite{Bakopoulos:2021liw, Bakopoulos:2022csr}. Finally, in the framework of the scalar tensor theories numerous solutions for 
solitons have been found \cite{Janis:1968zz,Fisher:1948yn,Wyman:1981bd, Agnese:1985xj, Roberts:1989sk,Kleihaus:2019rbg,Kleihaus:2020qwo, Faraoni:2021nhi,Baake:2021jzv,Herdeiro:2019iwl, Bakopoulos:2022csr}. The geometry of these local objects is characterized by the absence of singularities or horizons while asymptotically they are flat or (A)dS. Note also that in scalar-tensor theories there are solutions of ultra-compact black holes, which constitute a new kind of local solutions \cite{Bakopoulos:2020dfg}. These are regular compact black hole solutions  with horizon radius always smaller than  the  horizon  radius  of  the corresponding GR black hole with exactly the same mass. 
All of the above solutions may be accurate models for (ultra-)compact objects in the universe like the X-ray transient GROJ0422+32 \cite{compact} (see also \cite{Abadie:2010cf, LIGOScientific:2018mvr, Abbott:2020uma}).

In this work we will focus on a very simple action functional which besides gravity contains a minimally coupled scalar field with both its kinetic and potential terms.  This theory belongs to the class of the Horndeski theories,\,\footnote{It is a Horndeski theory with $G_2=X+V(\Phi)$, $G_4=1$ and $G_3=G_5=0$.} and it is conformally equivalent with the $f(R)$ and Brans-Dicke theory. The last decades, it is widely used in Cosmology since it provides accurate models for dark energy and inflation. Returning to the families of local solutions, the theory was used quite early on for the construction of wormhole solutions with the  most characteristic example the Ellis wormhole \cite{Ellis:1973yv, Ellis:1979bh,Bronnikov:1973fh}, which is supported by phantom fields. However, due to the No-Scalar hair theorem it is impossible to derive black hole solutions for a large class of potentials. It may be shown that black hole solutions may be found only for negative definite potentials, $V(\Phi)<0$. Also, although this theory is very simple, it is extremely difficult to provide analytic solutions for black holes. Even for the simple polynomial potentials $V(\Phi)=\lambda \Phi^n$ we need to employ numerical methods. Therefore, it is essential to map the potentials that lead to analytic solutions for black holes. The last years several works have been made towards this direction \cite{Bechmann:1995sa, Dennhardt:1996cz, Nucamendi:1995ex, Anabalon:2012ih, Cadoni:2015gfa}. These works have in common that in order to construct analytic solutions the authors employ the so-called ``scalar-potential engineering" method. In this, we specify a form for the scalar field and the potential is determined by solving the field equations. Here, we start from an analytic black hole solution that first appeared in \cite{Herdeiro:2015waa} and thoroughly study its properties. We show that these types of solutions can indeed describe ultra-compact black holes, we generalise these solutions into slow rotating ones, and we also compare their angular velocities with the angular velocities of the corresponding slow-rotating Schwarzschild black-hole solutions of the same mass. The most attractive feature of these solutions is that the scalar field assumes a coulombic form, while in order to deduce whether these types of solutions could comprise astrophysical objects, we also examine their thermodynamic stability and their stability under spacetime perturbations.

The outline of our work is as follows: In Section \ref{sec1}, we present our theoretical framework. We introduce our four-dimensional field theory and then we derive the black hole solution by solving the field equations. Furthermore, we study the properties of the solution, we determine the location of its horizon and we compare it with the horizon radius of the Schwarzschild  black  hole. Also, we examine the black-hole entropy, the regularity of the scalar curvature quantities and the energy conditions for our solution. Moreover, we investigate whether the potential that supports our solutions is in agreement with the conditions for the evasion of the No-scalar Hair theorem. In Section \ref{sec2}, we generalise the static solution into a slow-rotating one by using the slow rotation approximation. We begin by defining the necessary formalism for a general metric and then by substituting our solution we find the analytic expression of the angular velocity $\omega(r)$, which then is compared with the angular velocity of the corresponding slow-rotating Schwarzschild black hole. In Section \ref{sec3}, we consider the stability of our black hole solutions under linear perturbations. We derive the Schrödinger-like equation which is obeyed by the perturbations and by examining the effective potential we derive the stability conditions. Finally, in Section \ref{sec4} we conclude and discuss future directions.

\section{The Theoretical Framework}\label{sec1}

We consider the following action functional
\eq$\label{action}
\mathcal{S}=\frac{1}{16\pi}\int d^4x\sqrt{-g}\bigg[R-\frac{1}{2}(\pa \Phi)^2-V(\Phi)\bigg]\,.$
The theory contains the scalar curvature $R\equiv g^{\mu\nu}R_{\mu\nu}$, and a scalar field $\Phi$ with its kinetic\,\footnote{We note that $(\pa \Phi)^2\equiv \pa^\mu\Phi\pa_\mu\Phi$.} and potential terms. The gravitational field equations of the theory follow if we vary the above action with respect to the metric tensor $g_{\mu\nu}$, while the equation of motion of the scalar field follows from the variation of the action with respect to $\Phi$. By doing so, we obtain the following equations:
\gat$\label{gr-eqs}
G^{\mu}{}_\nu=T^{(\Phi)\mu}{}_\nu\,, \\[2mm]
\label{sc-eq}
\nabla^\lam \nabla_\lam \Phi-\pa_\Phi V =0\,,$
where $T^{(\Phi)\mu}{}_\nu$ is the stress-energy tensor associated with the scalar field $\Phi$ and is defined by
\eq$\label{stress-ten}
T^{(\Phi)\mu}{}_\nu\equiv \frac{1}{2}\pa^\mu\Phi\pa_\nu\Phi-\frac{1}{2}\del^{\mu}{}_\nu\left[\frac{(\pa \Phi)^2}{2}+V(\Phi)\right]\,.$

In the context of this work, we are interested in analytic black-hole solutions with scalar hair which are regular and asymptotically flat. To this end, we consider the geometric ansatz
\eq$\label{metr-ans}
ds^2=-e^{A(r)}\, B(r)\, dt^2+\frac{dr^2}{B(r)}+r^2\big(d\theta^2+\sin^2\theta\, d\varphi^2\big)\,,\hspace{1em} B(r)\equiv 1-\frac{2m(r)}{r}\,,$
while for the scalar field it is natural to assume only radial dependence, namely $\Phi=\Phi(r)$. Using now \eqref{gr-eqs} together with \eqref{stress-ten} we obtain the following independent equations
\gat$\label{eq1}
A'=\frac{r}{2} \left(\Phi'\right)^2\,,\\[3mm]
\label{eq2}
A''\big(r-2m\big)+A'\bigg(1+\frac{m}{r}-3m'\bigg)+\left(A'\right)^2\bigg(\frac{r}{2}-m\bigg)-2m''+\frac{4m'}{r}=0\,,\\[3mm]
\label{eq3}
V(\Phi)=\frac{2}{r}\bigg(\frac{2m}{r}-1\bigg)A'+\frac{4m'}{r^2}+\left(\Phi'\right)^2\bigg(\frac{1}{2}-\frac{m}{r}\bigg)\,.$
In the above, with prime we denote the derivative with respect to the radial coordinate $r$. Note also, that the scalar-field equation \eqref{sc-eq} is not an independent one, but rather follows from the aforementioned three differential equations. It is now obvious, that in order to solve the above set of differential equations \eqref{eq1}-\eqref{eq3} with four unknown functions, we first need to impose a particular expression for one of the functions involved. Consequently, assuming the scalar-field function $\Phi(r)=q/r$, where $q$ is the scalar charge, and solving one by one the above equations, we obtain
\eq$\label{A-r}
A(r)=-\frac{q^2}{4r^2}\,,$
\bal$\label{m-r}
m(r)=\frac{r}{2}+\frac{4r^3}{q^2}&+\frac{e^{\frac{q^2}{8r^2}}r^2}{q^2}\bigg[-12M+\sqrt{2\pi}\, q\, \erf\left(\frac{q}{2\sqrt{2}\,r}\right)\bigg]\nonum\\[1mm]
&-\frac{e^{\frac{q^2}{4r^2}}r^3}{q^3}\bigg\{4q-12\sqrt{2\pi}\, M\,\erf\left(\frac{q}{2\sqrt{2}\,r}\right)+\pi q \bigg[\erf\left(\frac{q}{2\sqrt{2}\,r}\right)\bigg]^2\bigg\}\,,$
\bal$\label{V-phi}
V(\Phi)=\frac{2(24+\Phi^2)}{q^2}&-\frac{12\,\Phi\, e^{\Phi^2/8}}{q^3}\bigg[12M-\sqrt{2\pi}\, q\, \erf\left(\frac{\Phi}{2\sqrt{2}}\right)\bigg]\nonum\\[1mm]
&+\frac{(\Phi^2-12)e^{\Phi^2/4}}{q^3}\bigg\{4q-12\sqrt{2\pi}\,M\, \erf\left(\frac{\Phi}{2\sqrt{2}}\right)+\pi q \bigg[\erf\left(\frac{\Phi}{2\sqrt{2}}\right)\bigg]^2\bigg\}\,,$
respectively.\,\footnote{The error function is defined as
$$\erf\, x=\frac{2}{\sqrt{\pi}}\int_{0}^x e^{-t^2}dt\,. $$} Notice that the scalar potential $V(\Phi)$ contains a term of the form $(2/q^2)\,\Phi^2$, therefore, the parameter $q$ can be related to the mass of the scalar field. In \eqref{m-r} we have chosen the integration constants appropriately in order to get a flat geometry at the spacetime boundary.\,\footnote{As mentioned in the Introduction, the above asymptotically flat solution has been first derived in the review paper \cite{Herdeiro:2015waa}, where the authors did not go into a detailed analysis of the solution, but rather mentioned it as an example of a black-hole solution with scalar hair.} Note that in general the integration constants may generate an effective cosmological constant $\Lambda_{\rm eff}$ resulting to an asymptotically (anti-)de Sitter spacetime. The general solution may be found in Appendix \ref{app: ads}. Here, we will focus on the asymptotically flat solutions, thus at infinity the metric function $A(r)$ gives
\eq$\label{A-asym}
\lim_{r\rightarrow +\infty} e^{A(r)}=1\,,$
while the expansion of $B(r)$ at large values of the $r-$coordinate results to
\eq$\label{B-asym}
B(r\gg 1)=1-\frac{2M}{r}+\frac{q^2}{4r^2}-\frac{7Mq^2}{20r^3}+\frac{q^4}{36r^4}+\mathcal{O}\left(\frac{1}{r^5}\right)\,.$
With the use of the above expressions in \eqref{metr-ans}, it is straightforward to confirm that at the radial infinity ($r\rightarrow +\infty$) the spacetime geometry is indeed flat. We also notice that the first three terms of the expansion resembles a Reissner-N\"ordstrom (RN) black hole, and since $e^{A(r)}$ goes to unity as $r$ grows larger, the spacetime geometry spanned by \eqref{metr-ans} cannot be distinguished from the geometry of a RS black hole at large values of the $r-$coordinate. However, in our case, the parameter $q$ originates from the scalar field $\Phi$ rather than from an electromagnetic gauge field. 

\begin{figure}[t]
    \centering
    \begin{subfigure}[b]{0.49\textwidth}
    \includegraphics[width=1\textwidth]{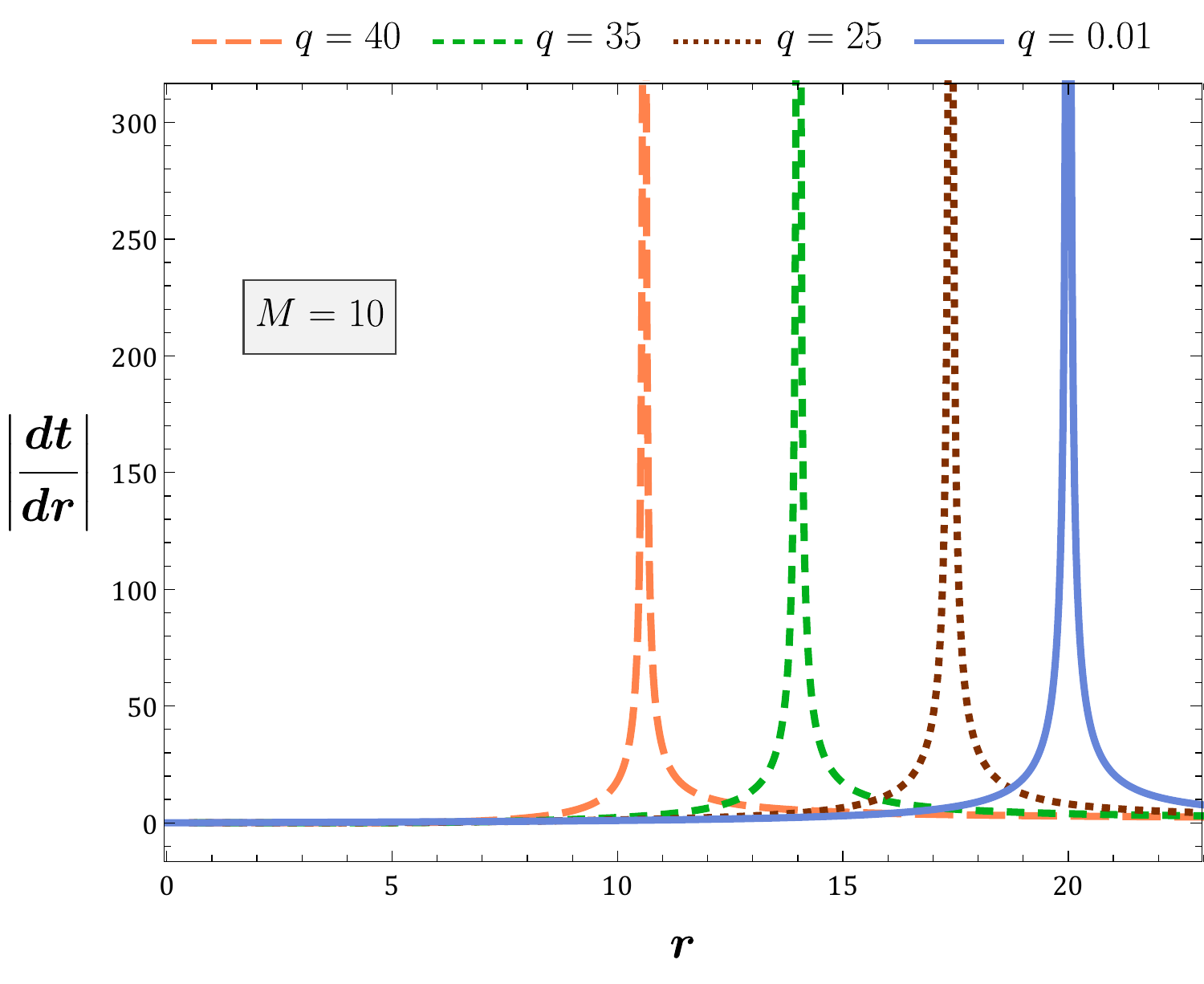}
    \caption{\hspace*{-3.5em}}
    \label{subf: horzs}
    \end{subfigure}
    \hfill
    \begin{subfigure}[b]{0.49\textwidth}
    \includegraphics[width=1\textwidth]{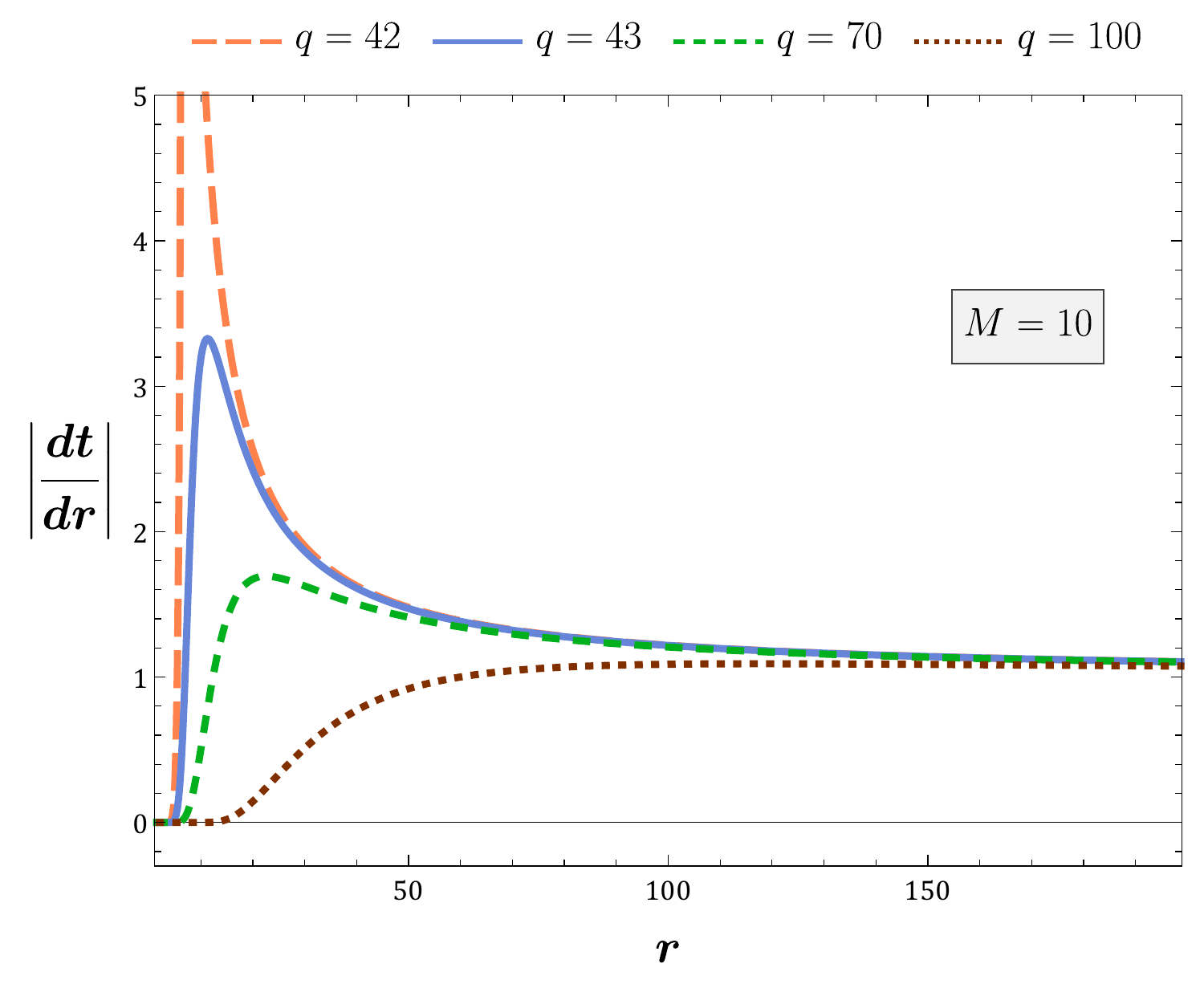}
    \caption{\hspace*{-3em}}
    \label{subf: naked}
    \end{subfigure}
    \caption{Both (a) and (b) depict graphs of the ratio $|dt/dr|$ in terms of the radial coordinate $r$ for various values of the parameter $q$ while keeping the mass parameter constant, i.e. $M=10$. In (a) $q=40,\, 35,\, 25,\, 0.01$ from left to right, while in (b) $q=42,\, 43,\, 70,\, 100$ from top to bottom.}
    \label{fig: null-tr}
\end{figure}

To obtain a better understanding of the spacetime geometry, we need to study its causal structure and the curvature invariant quantities generated by its line-element. In Appendix \ref{app: sc-curv}, we give the analytic expressions of all three scalar curvature quantities derived by (\ref{metr-ans}, \ref{A-r}, \ref{m-r}). Their expressions are fairly complex, but it is clear that at $r=0$ the spacetime exhibits a singularity, since all the terms of the form $1/r^{\ell}$ and $e^{q^2/r^2}$ diverge at $r=0$. We will come back to the scrutiny of the scalar curvature quantities shortly, but first it will prove helpful to investigate the horizon(s) of the black hole. The causal structure of the spacetime is defined via the light cone, and thus we need to consider radial null trajectories in the background geometry. Therefore, using \eqref{metr-ans} and by keeping the coordinates $\theta$ and $\varphi$ constant, the condition $ds^2=0$ leads to
\eq$\label{null-tr}
\frac{dt}{dr}=\pm \frac{e^{\frac{q^2}{8r^2}}}{|B(r)|}=\pm \frac{e^{\frac{q^2}{8r^2}}}{|1-\frac{2m(r)}{r}|}\,.$
At the spacetime boundary, namely when $r\rightarrow+\infty$, the above relation becomes $dt/dr=\pm 1$, as expected for asymptotically flat spacetimes. However, at the root(s) of the function $B(r)$ the fraction $dt/dr$ diverges. Therefore, the horizon radius $r_h$ is defined through the relation $B(r_h)=0$. In both Figs. \myref{fig: null-tr}{subf: horzs} and \myref{fig: null-tr}{subf: naked} we see the graphs of the positive branch of the quantity $dt/dr$ in terms of the radial coordinate $r$, for various values of the $q$-parameter and fixed mass value $M=10$. Since, there is at most one spike for each graph, we forthwith deduce that there is only one root for the relation $B(r_h)=0$, if any. Consequently, there is at most a single horizon for the black hole described by the line-element \eqref{metr-ans}. In Fig. \myref{fig: null-tr}{subf: horzs} for $q=0.01$ we have $r_h\simeq 20$, for $q=25$, $r_h\simeq 17.4$, for $q=35$, $r_h\simeq 14$, while for $q=40$, $r_h\simeq 10.6$. In Fig. \myref{fig: null-tr}{subf: naked} we observe that as the value of $q$ increases, there is a point ($q\gtrsim 42$) at which the graph smoothens and the spike disappears; this means that for these particular values of the parameters $M$ and $q$, the line-element \eqref{metr-ans} describes the topology of a naked singularity. Although we have obtained the above result for a particular value of the black-hole mass, namely $M=10$, one can verify that the transition from a black-hole solution to a naked singularity happens at the relative value $q/M\gtrsim 4.2$. This behaviour is also observed in the Reissner-Nordstr\"om black hole, where the naked singularity appears in the region of the parameter space for which $Q^2>M^2$.\,\footnote{In this case, the implied Reissner-Nordstr\"om line-element is of the following form:
$$ds^2=-\left(1-\frac{2M}{r}+\frac{Q^2}{r^2}\right)dt^2+\left(1-\frac{2M}{r}+\frac{Q^2}{r^2}\right)^{-1}dr^2+r^2\,
d\Omega^2_2\,. $$} Of course, we do not expect these particular paradigms of naked singularities to correspond to astrophysical objects.

\begin{figure}[t]
    \centering
    \includegraphics[width=0.55\textwidth]{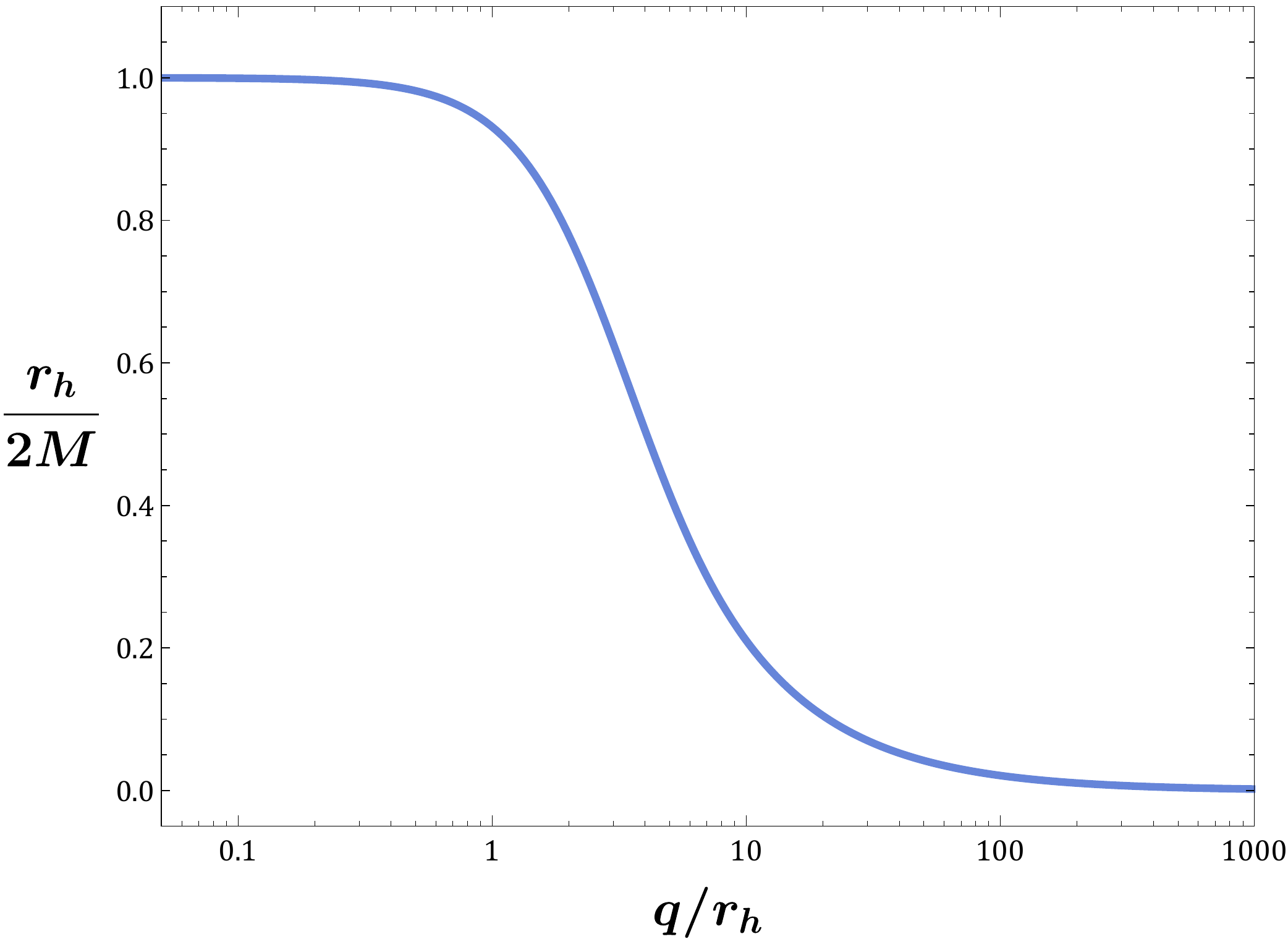}
    \caption{The graph of the fraction $r_h/(2M)$ in terms of $q/r_h$. Both quantities are dimensionless, while the horizontal axis is logarithmic.}
    \label{fig: UCBH}
\end{figure}

As we already discussed, from equation $B(r_h)=0$ one can determine the horizon radius $r_h$ in terms of the mass parameter $M$ and the parameter $q$. However, this value can be specified only numerically, since it is impossible to solve explicitly the equation $B(r_h)=0$ with respect to $r_h$. It is though possible and even more meaningful to solve the aforementioned equation with respect to the dimensionless quantity $r_h/(2M)$. Using \eqref{m-r} and after a little bit of algebra it is easy to show that
\eq$\label{rh/M}
\frac{r_h}{2M}=\frac{6\,e^{\frac{q^2}{8 r_h^2}} \left[\displaystyle{\frac{q}{r_h}}-\sqrt{2 \pi }\, e^{\frac{q^2}{8 r_h^2}}\, \text{erf}\left(\frac{q}{2 \sqrt{2}\, r_h}\right)\right]}{\displaystyle{\frac{q}{r_h}} \left\{4-4\, e^{\frac{q^2}{4 r_h^2}}+\frac{q}{r_h} \sqrt{2 \pi }\, e^{\frac{q^2}{8 r_h^2}}\, \text{erf}\left(\frac{q}{2 \sqrt{2}\, r_h}\right)-\pi \,e^{\frac{q^2}{4 r_h^2}} \left[\text{erf}\left(\frac{q}{2 \sqrt{2}\, r_h}\right)\right]^2\right\}}\,.$
From \eqref{rh/M}, it is obvious that the value of the quantity $r_h/(2M)$ depends only on the value of $q/r_h$, therefore the scalar hair is of secondary type. In Fig. \ref{fig: UCBH} we see the graph of the above relation. Notice that for small values of the ratio $q/r_h$,\,\footnote{The expansion of the quantity $r_h/(2M)$ for $q/r_h\ll 1$ is the following: $$\frac{r_h}{2M}=1-\frac{3}{40}\left(\frac{q}{r_h}\right)^2
+\frac{55}{8064}\left(\frac{q}{r_h}\right)^4+\mathcal{O}\bigg[\bigg(\frac{q}{r_h}\bigg)^6\bigg]\,.$$} the fraction $r_h/(2M)$ is equal to unity and therefore $r_h=2M$ as in the Schwarzschild geometry. We also observe that the greater the value of $q/r_h$, the smaller the value of $r_h/(2M)$; thus, in the region of the parameter space where $q>r_h$ the condition $B(r_h)=0$ leads to an ultra-compact black hole, since for a fixed value of the mass $M$ the horizon radius $r_h$ is smaller than the horizon radius of the corresponding Schwarzschild black hole with the same mass. Of course, for values $q/r_h>10$ the horizon radius of the ultra-compact black hole becomes extremely small and hence, it is very likely that this region of the parameter space is unphysical. Observational measurements from LIGO-VIRGO, Event Horizon Telescope or other astrophysical experiments/missions will prove very useful in the future and hopefully will put bounds on the size and mass of (ultra-)compact objects \cite{Abadie:2010cf, LIGOScientific:2018mvr, Abbott:2020uma}.

The horizon radius is also related to the entropy of the black hole. The entropy may be calculated using Wald's formula \cite{Wald:1993nt, Iyer:1994ys} that associates the Noether charge on the horizon with the entropy of the black hole
\begin{equation}
    S=-2\pi\oint d^2 x \sqrt{h_2}\left( \frac{\partial \mathcal{L}}{\partial R_{abcd}} \right)_{\mathcal{H}}\hat{\epsilon}_{ab}\,\hat{\epsilon}_{cd}\,.
\end{equation}
In the above, $\hat{\epsilon}_{ab}$ is the binormal to surface of the horizon $\mathcal{H}$, $h_2$ the determinant of the 2-dimensional projected metric on $\mathcal{H}$, and $\mathcal{L}$ the Lagrangian of the theory. By using the above equation it is easy to verify that in our theory the entropy is described by the Bekenstein-Hawking formula $S=\mathcal{A}/4$. Therefore, 
\begin{equation}
    \frac{S}{S_{\rm Sch}}=\left( \frac{r_h}{2M}\right)^2,
\end{equation}
where $S_{\rm Sch}$ is the entropy of the corresponding Schwarzschild black hole with mass $M$. For more details about the calculation of the entropy see \cite{Bakopoulos:2018nui}. From Fig. \ref{fig: UCBH}  we observe that black holes with $q/r_h<1$ have approximately the same entropy with the corresponding Schwarzschild black hole of the same mass, while the ultra-compact black holes have always smaller entropy than the Schwarzschild ones. Therefore, we expect the solutions with $q/r_h<1$ to be thermodynamically more stable than the ultra-compact black holes.

To comprehend fully the geometry of the spacetime (\ref{metr-ans}, \ref{A-r}, \ref{m-r}), we need to examine the curvature invariant quantities: the Ricci scalar $R$, the scalar $\mathcal{R}\equiv R^{\mu\nu}R_{\mu\nu}$, and the Kretschmann scalar $\mathcal{K}\equiv R^{\mu\nu\kappa\lam}R_{\mu\nu\kappa\lam}$. Their analytic expressions are presented in Appendix \ref{app: sc-curv}, while in Figs. \myref{fig: sc-curv}{subf: Ricci}, \myref{fig: sc-curv}{subf: Ricci2} and \myref{fig: sc-curv}{subf: Riem} we depict their graphs in terms of the dimensionless radial quantity $r/r_h$ for various values of the dimensionless parameter $q/r_h$. In all sub-figures of Fig. \ref{fig: sc-curv}, we have multiplied each scalar with the appropriate power of $r_h$ to make the resulting quantity dimensionless.\,\footnote{Note that $[R]=[L]^{-2}$, $[\mathcal{R}]=[\mathcal{K}]=[L]^{-4}$.} In addition, in order to keep the black-hole horizon constant in all graphs, independently of the value of $q$, we have substituted the mass $M$ from \eqref{rh/M} inside the relations \eqref{Ricci}-\eqref{Riem-sc}. By doing so, although we vary the parameter $q$, as long as we vary accordingly the mass of the black hole we can keep the black-hole horizon fixed. As one can clearly observe from Fig. \ref{fig: sc-curv}, all three scalar curvature quantities diverge at $r=0$. Therefore, we deduce that at $r=0$ we encounter a real spacetime singularity. However, as we move away from the black-hole singularity, the curvature invariant quantities obtain their zero asymptotic values extremely fast, hence leading to an asymptotically flat spacetime. One can also notice that for values $q\lesssim r_h$ the scalar quantities are effectively zero on the black-hole horizon, and thus, the intense change of the spacetime curvature due to the existence of the black-hole singularity at $r=0$ is only noticeable way beyond the crossing of the horizon, namely at distance $r\simeq r_h/2$. On the other hand, for values $q\geq 5r_h$, the effect of the singularity on the curvature of the spacetime becomes apparent even to an observer outside the black-hole horizon. This behaviour is indeed anticipated and in complete agreement with the discussion which took place earlier, since as it was shown in Fig. \ref{fig: UCBH} for values $q/r_h\geq 5$ we have an ultra-compact black hole.

\begin{figure}[t]
    \centering
    \begin{subfigure}[b]{0.49\textwidth}
    \includegraphics[width=1\textwidth]{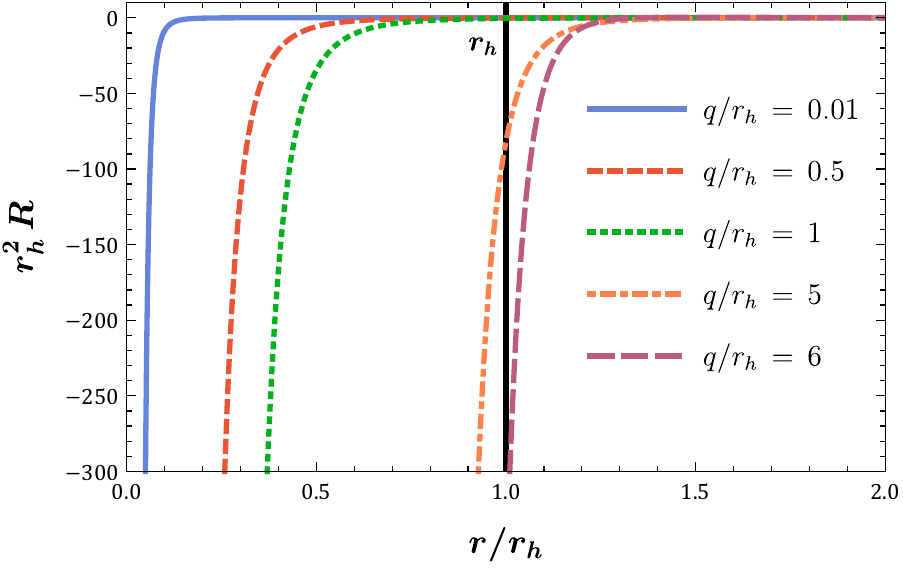}
    \caption{\hspace*{-3.3em}}
    \label{subf: Ricci}
    \end{subfigure}
    \hfill
    \begin{subfigure}[b]{0.5\textwidth}
    \includegraphics[width=1\textwidth]{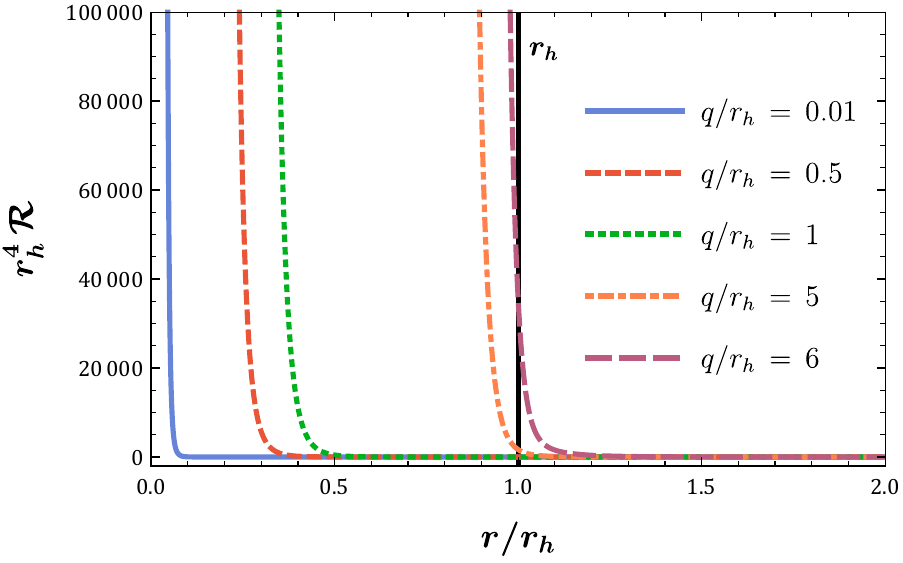}
    \caption{\hspace*{-4em}}
    \label{subf: Ricci2}
    \end{subfigure}
    \begin{subfigure}[b]{0.5\textwidth}
    \vspace*{1em}
    \includegraphics[width=1\textwidth]{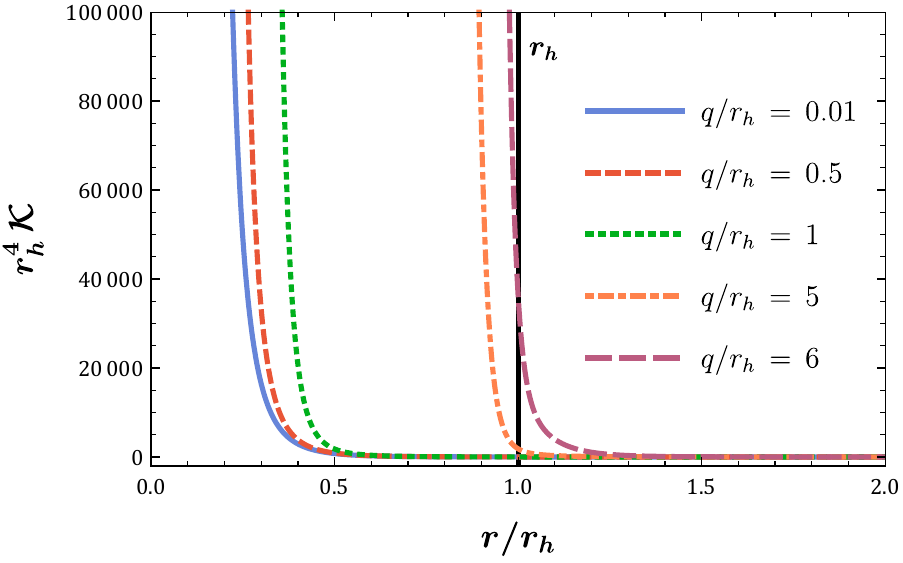}
    \caption{\hspace*{-4.2em}}
    \label{subf: Riem}
    \end{subfigure}
    \caption{(a) The Ricci scalar $r_h^2\,R$, (b) the scalar $r_h^4\, \mathcal{R}$ and (c) the Kretschmann scalar $r_h^4\,\mathcal{K}$ in terms of the radial coordinate $r$ for various values of the dimensionless parameter $q/r_h$. The quantity $q/r_h$ takes the values $0.01,\, 0.5,\, 1,\, 5,\, 6$ (from left to right).}
    \label{fig: sc-curv}
\end{figure}

\begin{figure}[t]
    \centering
    \begin{subfigure}[b]{0.49\textwidth}
    \includegraphics[width=1\textwidth]{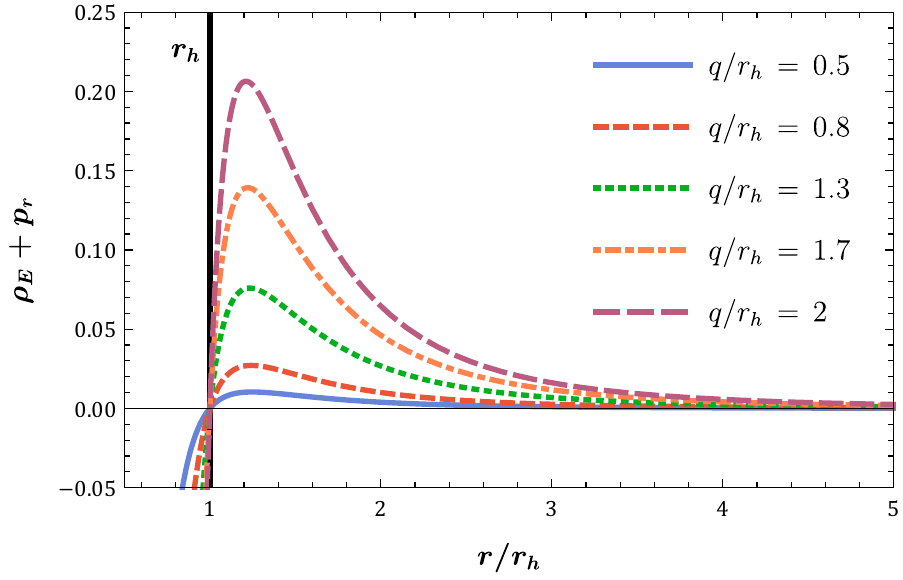}
    \caption{\hspace*{-3.5em}}
    \label{subf: SEC1}
    \end{subfigure}
    \hfill
    \begin{subfigure}[b]{0.5\textwidth}
    \includegraphics[width=1\textwidth]{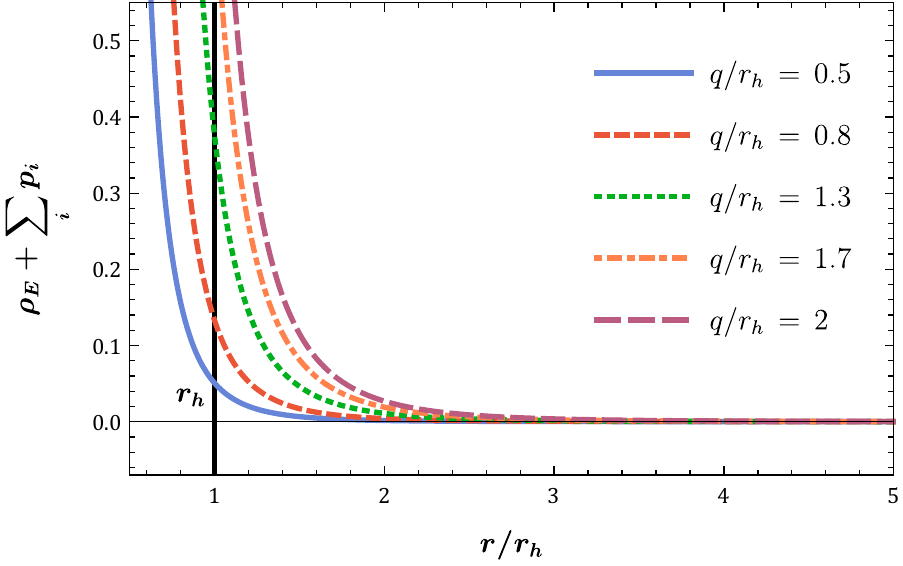}
    \caption{\hspace*{-3.7em}}
    \label{subf: SEC2}
    \end{subfigure}
    \caption{(a) The quantity $\rho_E+p_r$ and (b) the quantity $\rho_E+\sum_i p_i$ in terms of the radial coordinate $r$ for $r_h=1$
    and different values of the dimensionless parameter $q/r_h$.}
    \label{fig: SEC}
\end{figure}

Having studied the geometrical characteristics of the line-element (\ref{metr-ans}, \ref{A-r}, \ref{m-r}) it is now important to examine the energy conditions which are satisfied by the stress-energy tensor regarding the scalar field, namely $T^{(\Phi)\mu}{}_\nu$. It is easily deducible from \eqref{stress-ten} that the stress-energy tensor is solely characterized by three components: the energy density $\rho_E=-T^{(\Phi)t}{}_t$, the radial pressure $p_r=T^{(\Phi)r}{}_r$ and the tangential pressure $p_\theta=T^{(\Phi)\theta}{}_\theta=T^{(\Phi)\varphi}{}_\varphi$. It is also discernible that $p_\theta=-\rho_E$, while $p_r=w_r\, \rho_E$ with $w_r=w_r(r)$. In Fig. \myref{fig: SEC}{subf: SEC1} and \myref{fig: SEC}{subf: SEC2} we see the graphs of the quantities $\rho_E+p_r$ and $\rho_E+\sum_i p_i \equiv \rho_E + p_r + 2p_\theta = p_r + p_\theta$, respectively, in terms of the radial coordinate $r$, for $r_h=1$ and various values of the dimensionless parameter $q/r_h$. Note here that the mass $M$ inside the expression of the scalar potential $V(\Phi)$ and consequently inside the expressions of the energy density $\rho_E$ and the pressures $p_r$, $p_\theta$ has been replaced from Eq. \eqref{rh/M}. That is the reason why the only independent parameters of the quantities depicted in Fig. \ref{fig: SEC} are the horizon radius $r_h$ and the scalar charge $q$. One can readily observe that independently of the values of the ratio $q/r_h$, the quantity $\rho_E+\sum_i p_i$ is positive-definite for all values of the radial coordinate $r$, while $\rho_E+p_r$ is positive-definite for all values of $r$ which are greater than the black-hole horizon radius $r_h$, i.e. in the spacetime region which lies outside of the black hole. In Appendix \ref{app: ene-mom} one may find the explicit calculation which proves that the quantity $\rho_E+p_r$ changes sign at $r=r_h$, a result which is independent of the value of the parameter $q/r_h$. Therefore, combining the results from Figs. \myref{fig: SEC}{subf: SEC1} and \myref{fig: SEC}{subf: SEC2} we conclude that outside of the black hole ($r>r_h$)---in the causal region of spacetime---the distribution of energy and pressure corresponding to the scalar field $\Phi$ satisfies the strong energy conditions (SEC): $\rho_E+p_i\geq 0,\, \forall p_i $ and $\rho_E+\sum_{i}p_i\geq 0$.\,\footnote{For a perfect fluid $\rho+\sum_i p_i=\rho+3p$, and the second condition would have been $\rho+3p\geq 0$.} In the region of spacetime beyond the black hole horizon (i.e. for $r<r_h$), both the energy density $\rho_E$ and the quantity $\rho_E+p_r$ become negative, thus the energy conditions are violated. In wormhole solutions, we often meet violation of the energy conditions \cite{Morris:1988cz,Visser:1995cc,Kanti:2011jz,Kanti:2011yv,Antoniou:2019awm}, but 
for black hole solutions this type of matter is rarely encountered. In this particular scenario though, we see that for the interior of the black hole this type of exotic behaviour of the scalar field is indeed necessary, otherwise the simple scalar-field theory $\mathcal{L}_\Phi=-(\pa \Phi)^2/2-V(\Phi)$ would not have been able to support by itself the geometry of the ultra-compact black-hole solution. 

\begin{figure}[t]
    \centering
    \includegraphics[width=0.55\textwidth]{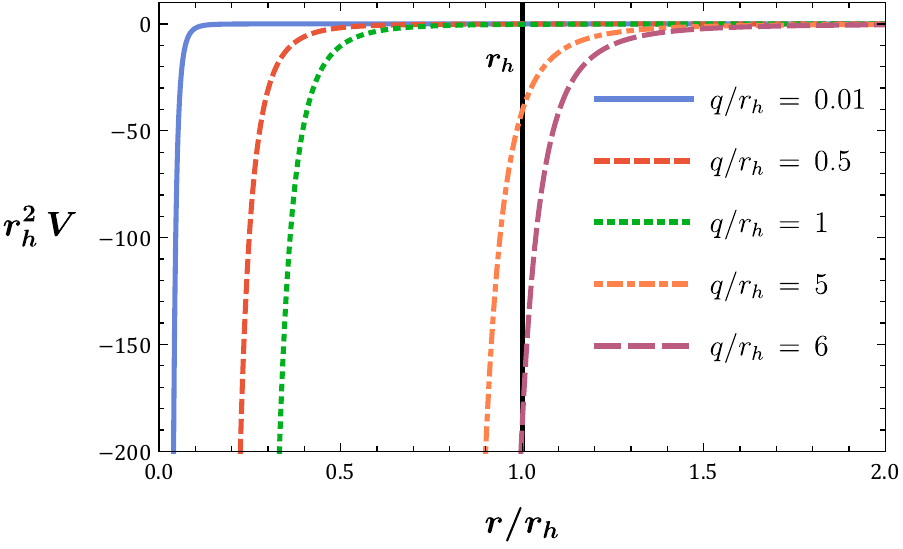}
    \caption{The potential $r_h^2\,V(\Phi(r))$ in terms of the radial coordinate $r$ for different values of the dimensionless parameter $q/r_h$.}
    \label{fig: pot}
\end{figure}

Before we discuss the scalar potential $V(\Phi)$ given by Eq. (\ref{V-phi}), let us first focus on the No-Scalar Hair theorem. The No-Scalar hair theorem  that has been developed by Bekenstein and Teitelboim \cite{NH-scalar1,NH-scalar2,Bekenstein}, is an extension of the GR No-Hair theorems which forbid the association of the black holes with scalar hair.  For the particular  theory that we consider in this work, the validity of the No-Scalar Hair theorem relies on the sign of the potential. We may easily verify this by multiplying the scalar field equation (\ref{sc-eq}) with $V$ and integrating over the whole spacetime outside the black hole
\begin{equation}
    \int_{r_h}^\infty dx^4 \sqrt{-g}\, V\left( \nabla_\mu\nabla^\mu \Phi-\pa_\Phi V \right)=0\,.
\end{equation}
Here we assume that the black hole is asymptotically flat and the scalar field has the same symmetries with the spacetime i.e. $\Phi(x^\mu)=\Phi(r)$. 
Integrating by parts the first term, the above equation becomes 
\begin{equation}
   \int_{r_h}^\infty dx^4 \sqrt{-g}\, \pa_\Phi V\left( \partial_\mu \Phi\,\partial^\mu \Phi+ V \right)=0\,.  \label{noh2}
\end{equation}
The boundary term $\left[\sqrt{-g}\,V\partial^\mu\Phi \right]_{r_h}^\infty$ vanishes in both boundaries. Near the horizon of the black hole it vanishes due to the factor $B(r)$ that appears via the derivative of the scalar field with respect to $r$,\,\footnote{$\partial^\mu\Phi=\del^{\mu r}g^{rr}\pa_r\Phi=\del^{\mu r}B(r)\pa_r\Phi$ and $B(r_h)=0$ by definition.}  while at infinity the scalar potential $V(\Phi)$ vanishes since the black-hole solution is asymptotically flat. Moreover, because of the staticity and spherical symmetry of our solutions, the first term in Eq. (\ref{noh2}) is $\partial_\mu \Phi\, \partial^\mu \Phi=g^{rr}\Phi'^2>0$ throughout the exterior region of the black hole. Therefore Eq. (\ref{noh2}) may be satisfied only if the potential is negative $V(\Phi)<0$. Hence, for a large class of theories with $V>0$, including the mass term $V=m^2 \Phi^2/2$, it is not possible to derive a non-trivial black hole solution. In these cases, the only acceptable spherically symmetric solution is the Schwarzschild one accompanied with a trivial (constant everywhere) scalar field. Returning to  our solutions, in Fig. \ref{fig: pot} we depict the potential $r_h^2\,V(\Phi(r))$ for five black hole solutions with different values of the dimensionless parameter $q/r_h$. We have multiplied the potential with $r_h^2$ in order to make the resulting quantity dimensionless and also scale invariant, as long as we adjust accordingly the mass of the black hole. We observe that in accordance with the evasion condition, the black-hole solutions are always accompanied by a negative potential. For $q/r_h<1$ the effect of the potential is significant only inside the black hole while it is negligible at the horizon. On the other hand, the potential of the ultra-compact black holes is also considerable in the near horizon region. For every solution, the potential always obeys the condition $V(\Phi)<0$, which means that is negative and has no nodes. Finally, at infinity, the potential vanishes and this leads to the asymptotic flatness of our solutions.

\section{Black-hole solution with slow rotation}\label{sec2}

In this section, we will generalise the static line-element \eqref{metr-ans} in order to incorporate black-hole solutions with slow rotation. To this end, we will approximate the rotating solutions as axisymmetric perturbations on the static and spherically symmetric spacetime \eqref{metr-ans}. Following the same method as Pani and Cardoso did in \cite{Pani:2009wy}, which constitutes a
generalization of the Hartle's method \cite{Hartle:1967he} in the framework of General Relativity, we are led to consider the line-element
\eq$\label{slow-metr}
{ds}^2=-e^{A(r)}B(r)\,{dt}^2+\frac{dr^2}{B(r)}+r^2\left\{{d\theta}^2+\sin^2\theta\left[d\varphi-\varepsilon\, \Omega(r,\theta)\, dt\right]^2\right\}.$
In the above, $\Omega(r,\theta)$ is the angular velocity of an observer at $(r,\theta)$, while $\varepsilon$ is a dimensionless parameter which helps us control the perturbations. We also assume that the scalar field is $\Phi^{tot}(r,\theta)=\Phi(r)+\varepsilon\, \Phi_1(r,\theta)$.  We take the expansion at $\varepsilon\rightarrow 0$ and we find that the unperturbed system of equations accept the solution described by Eqs. (\ref{A-r}-\ref{V-phi}) of the previous section. The independent equations to first order in $\varepsilon$, namely $\mathcal{O}(\varepsilon)$,
have the form
\gat$
(\pa_\Phi V)\Phi_1(r,\theta)+B(r)(\pa_r\Phi)\partial_r\Phi_1(r,\theta)=0\,,\label{phir1}\\[2mm]
\partial_\theta\Phi_1(r,\theta)=0\,, \label{phir2}\\[2mm]
3\cot \theta\,\partial_\theta\Omega(r,\theta)+\partial_\theta^2\Omega(r,\theta)+\frac{rB(r)}{2}(8-r\pa_r A)\,\partial_r\Omega(r,\theta)+r^2B(r)\,\partial_r^2\Omega(r,\theta)=0\,.\label{pert-eq}  
$
Equation (\ref{phir2}) indicates that the scalar correction $\Phi_1$ has no angular dependence, hence $\Phi_1(r,\theta)=\Phi_1(r)$.
Using the previous result in eq. (\ref{phir1}) we find that
\eq$\label{phi1-expr}
\Phi_1(r)=\Phi_0\, \exp\left(-\int \frac{\pa_\Phi V}{B(r)\pa_r\Phi}\, dr\right)\,,$
where $\Phi_0$ is an integration constant. 
Although the complexity of the functions $V(\Phi)$ and $B(r)$ does not allow us to explicitly compute the integral in eq. \eqref{phi1-expr}, we
are able to calculate it near the black-hole horizon and asymptotically.
In the asymptotic limit the expression inside the integral of \eqref{phi1-expr} takes the form
\eq$
\frac{\pa_\Phi V}{B(r)\pa_r\Phi}\xrightarrow{r\rightarrow+\infty}\frac{2M}{r^2}+\frac{16M^2-q^2}{4r^3}+\mathcal{O}\left(\frac{1}{r^4}\right)\,,$
which by its turn leads to 
\eq$\label{phi1-asym}
\Phi_1(r)\xrightarrow{r\rightarrow+\infty}\Phi_0\left[1+\frac{2M}{r}+\frac{16M^2-q^2}{8r^2}+\mathcal{O}\left(\frac{1}{r^3}\right)\right]\,.$
On the other extreme, that is near the black-hole horizon, and by using eq. \eqref{rh/M} one finds that
\eq$
\frac{\pa_\Phi V}{B(r)\pa_r\Phi}\xrightarrow{r\rightarrow r_h}\frac{1}{r-r_h}+\mathcal{A}+\mathcal{O}(r-r_h)\,,$
where $\mathcal{A}=\mathcal{A}(r_h,q)$ and is given by
{\fontsize{10}{10}{\eq$
\mathcal{A}=\frac{e^{-\frac{q^2}{8 r_h^2}} \left[4 q \left(q^2 e^{\frac{q^2}{4 r_h^2}}-4 r_h^2\right)-4 \sqrt{2 \pi }\, r_h\, e^{\frac{q^2}{8 r_h^2}} \left(q^2-4 r_h^2\right) \text{erf}\left(\frac{q}{2 \sqrt{2} r_h}\right)-e^{\frac{q^2}{4 r_h^2}}\pi  q^3 \text{erf}^{\,2}\left(\frac{q}{2 \sqrt{2} r_h}\right)\right]}{8 r_h^3 \left[\pi  q\, e^{\frac{q^2}{8 r_h^2}} \text{erf}^{\,2}\left(\frac{q}{2 \sqrt{2} r_h}\right)+4 \sqrt{2 \pi }\, r_h\, \text{erf}\left(\frac{q}{2 \sqrt{2}
   r_h}\right)-4 q e^{\frac{q^2}{8 r_h^2}}\right]}\,.$}}
Consequently, the near horizon expansion results to
\eq$\label{phi1-hor}
\Phi_1(r)\xrightarrow{r\rightarrow r_h}\Phi_0\, \frac{e^{-\mathcal{A}\, r}}{r-r_h}\,.$
We observe that $\Phi_1(r)$ diverges at the horizon and therefore we are obliged to set $\Phi_0=0$, otherwise the slow-rotating approximation does not hold. This means that for the assumed Lagrangian density, and metric ansatz, the slow-rotating approximation demands a scalar field which is of purely coulombic form. Of course, if one tries to find the non-perturbative generalization of the line-element \eqref{metr-ans}, then it is indisputable that the scalar field would differ significantly from the coulombic form. However, this is not the case here. 

Let us now turn to eq. \eqref{pert-eq}. By using the method of separation of variables, eq. (\ref{pert-eq}) can be decomposed into two differential equations, one for each coordinate 
\gat$
\label{pert-ang}
3\cot\theta\frac{dY_\ell}{d\theta}+\frac{d^2Y_\ell}{d\theta^2}=-\ell(\ell+1)Y_\ell\,,\\[2mm]
\label{pert-r}
r^2B\,\omega_\ell''+\frac{rB}{2}(8-rA')\,\omega_\ell'=\ell(\ell+1)\,\omega_\ell\,.$
Therefore, the general solution of Eq. \eqref{pert-eq} is $\Omega(r,\theta)=\sum_\ell\omega_\ell(r)Y_\ell(\theta)$.
The slow rotation approximation of the Schwarzschild solution gives
\eq$\label{omega-Sch}
\omega_{\rm Sch}(r)=\frac{2 J}{r^3}\,,$
where $J$ is the angular momentum of the black hole \cite{Pani:2009wy}. Consequently, the above relation will serve as a boundary condition for \eqref{pert-r}. At infinity we find that \eqref{pert-r} takes the form
\begin{align}
\omega_\ell''+\frac{4}{r}\omega_\ell'-\frac{\ell(\ell+1)}{r^2}\omega_\ell=0\,,
\end{align}
with its solution being
\begin{align}
\omega_\ell(r)=c_1\,r^{b_{+}}+c_2\,r^{b_-}, \quad \text{with} \quad b_{\pm}=-\frac{1}{2}\left( 3\pm\sqrt{9+4\ell(\ell+1)}\,\, \right).
\end{align}
We observe that only the solution with $\ell=0$, $c_1=2J$ and $c_2=0$ satisfies the boundary condition \eqref{omega-Sch}. For $\ell=0$, Eq. (\ref{pert-ang}) assumes the solution $Y_0(\theta)=const$, which means that the general solution of the angular velocity reduces to
\begin{align}
\Omega(r,\theta)=\Omega(r)=\omega(r)\,.
\end{align}
Thus, we conclude that at first order in perturbation theory, the solution for a slowly rotating black hole does not have angular dependence. Hence, Eq. \eqref{pert-r} leads to the following differential equation for $\omega(r)$:
\begin{align}
\label{omega-dif}
\omega''+\frac{1}{2r}\left(8-\frac{q^2}{2r^2}\right)\omega'=0\,.
\end{align}
The solution of the above differential equation is
\eq$\label{omega-r}
\omega(r)=-\frac{24J}{q^3} \left[\frac{q\, e^{-\frac{q^2}{8 r^2}}}{r}-\sqrt{2 \pi }\, \text{erf}\left(\frac{q}{2 \sqrt{2}\, r}\right)\right],$
where we have already taken into account the appropriate asymptotic behaviour of $\omega(r)$, and we have identified the integration constants by matching the asymptotic expansion\,\footnote{For large values of the radial coordinate, $r\gg 1$, it holds that $$\omega(r)=\frac{2J}{r^3} - \frac{3q^2J}{20\,r^5} + \frac{3q^4J}{448\, r^7} + \mathcal{O}\left(\frac{1}{r^9}\right) \simeq \frac{2J}{r^3}\,.$$} of $\omega(r)$ with $\omega_{\rm Sch}(r)$.

\begin{figure}[t]
    \centering
    \includegraphics[width=0.55\textwidth]{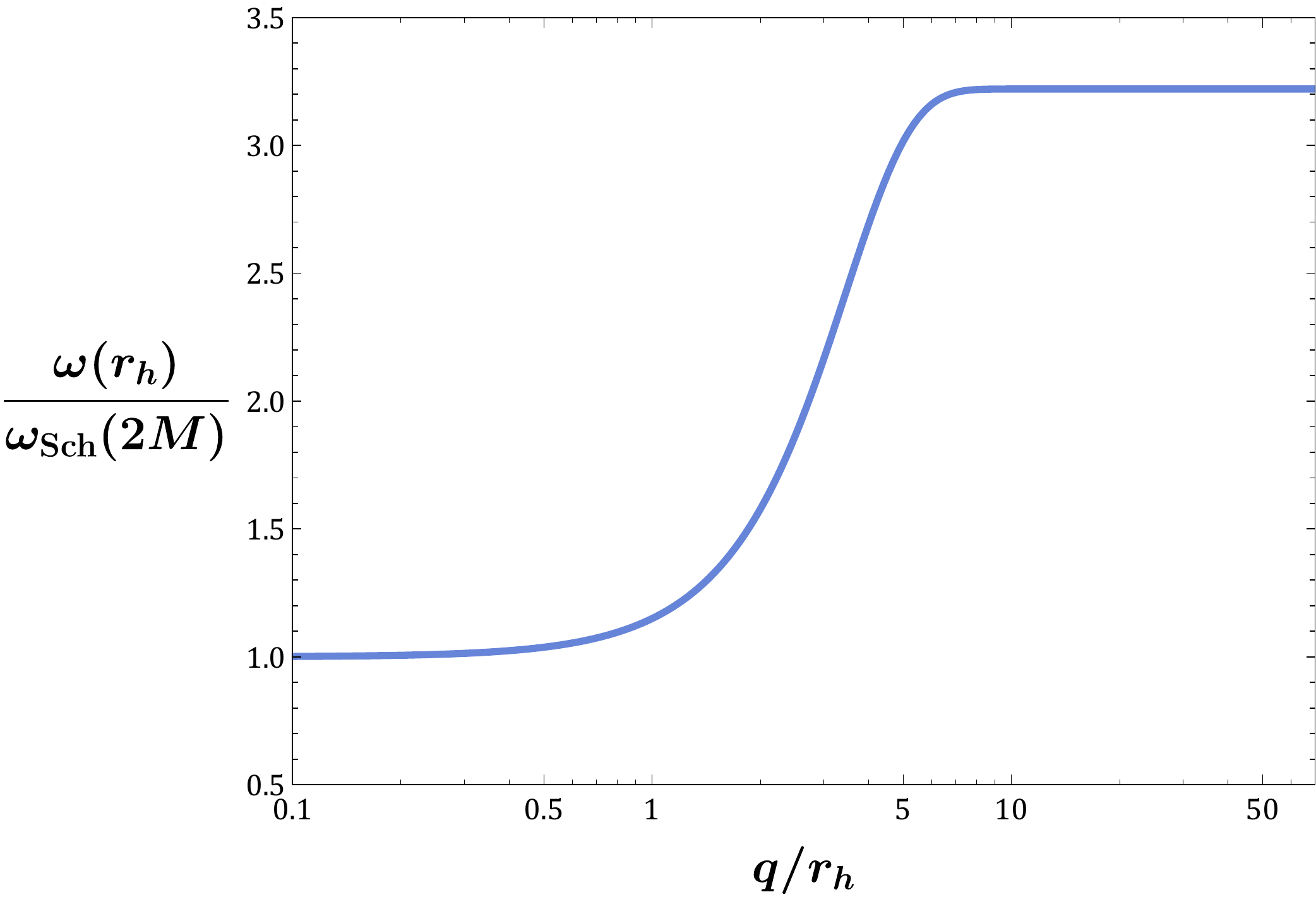}
    \caption{The graph of the ratio $\omega(r_h)/\omega_{\rm Sch}(2M)$ in terms of $q/r_h$. Both quantities are dimensionless, while the horizontal axis is logarithmic.}
    \label{fig: Ang}
\end{figure}

Summing up, we have shown that the action functional \eqref{action} with $\Phi(r)=q/r$ and $V(\Phi)$ given by the relation \eqref{V-phi}, incorporates as a solution a slow rotating ultra-compact black hole described by the line-element
\eq$\label{slow-rot-UCBH}
{ds}^2=-e^{A(r)}B(r)\,{dt}^2+\frac{dr^2}{B(r)}+r^2\left\{{d\theta}^2+\sin^2\theta\left[d\varphi-\varepsilon\, \omega(r)\, dt\right]^2\right\},\hspace{1em} B(r)=1-\frac{2m(r)}{r}\,.$
In the above, the functions $A(r)$, $m(r)$ and $\omega(r)$ are given by Eqs. \eqref{A-r}, \eqref{m-r} and \eqref{omega-r}, respectively.

Considering now the slow-rotating ultra-compact black hole \eqref{slow-rot-UCBH} and a slow-rotating Schwarzschild black hole of the same mass, we can calculate the ratio of their angular velocities evaluated respectively at each horizon: $\omega(r_h)/\omega_{\rm Sch}(2M)$. With the use of Eqs. \eqref{omega-r}, \eqref{omega-Sch} and \eqref{rh/M}, we obtain the following expression
\eq$\label{omega-R}
\frac{\omega(r_h)}{\omega_{\rm Sch}(2M)}=-\frac{\left\{4-4\, e^{\frac{q^2}{4 r_h^2}}+\displaystyle{\frac{q}{r_h}} \sqrt{2 \pi }\, e^{\frac{q^2}{8 r_h^2}}\, \text{erf}\left(\frac{q}{2 \sqrt{2}\, r_h}\right)-\pi \,e^{\frac{q^2}{4 r_h^2}} \left[\text{erf}\left(\frac{q}{2 \sqrt{2}\, r_h}\right)\right]^2\right\}^3}{18\,e^{\frac{3q^2}{4 r_h^2}}\left[\displaystyle{\frac{q}{r_h}}\, e^{-\frac{q^2}{8 r_h^2}}-\sqrt{2 \pi }\, \text{erf}\left(\frac{q}{2 \sqrt{2}\, r_h}\right)\right]^2}\,.$
In Fig. \ref{fig: Ang}, we depict the graph of the ratio $\omega(r_h)/\omega_{\rm Sch}(2M)$ in terms of the dimensionless quantity $q/r_h$. As mentioned above, we consider the case in which these two compact objects are of the same mass $M$, however, as we have already shown in Fig. \ref{fig: UCBH}, depending on the value of the parameter $q/r_h$ their horizon radii can differ significantly. We observe that for small values of the quantity $q/r_h$ the angular velocities have the same value. This is indeed anticipated by the fact that for small values of the parameter $q/r_h$, the horizon radius of the black-hole solution (\ref{metr-ans}, \ref{A-r}, \ref{m-r}) is equal to the horizon radius of the corresponding Schwarzschild one with the same mass, namely $r_h=2M$ (see Fig. \ref{fig: UCBH}). 
We also notice that the angular velocities remain fairly equal up until the value $q/r_h\simeq 1$, while from $q/r_h\simeq 1$ to $q/r_h\simeq 7$ the angular velocity $\omega(r_h)$ triples in magnitude compared to $\omega_{\rm Sch}(2M)$. After the point where $q/r_h\simeq 7$, the relative value of the angular velocities remains constant.
 
\section{Stability analysis}\label{sec3}

In this section, we will investigate the linear stability of the black-hole solution (\ref{metr-ans}, \ref{A-r}, \ref{m-r}) under small perturbations $h_{\mu\nu}$ in the background spacetime $g_{\mu\nu}$, $|h_{\mu\nu}|\ll |g_{\mu\nu}|$. The sum of the background metric $g_{\mu\nu}$ together with the spacetime perturbations $h_{\mu\nu}$ will constitute from now on the total metric tensor $g^{tot}_{\mu\nu}$:
\eq$\label{metr-tot}
g^{tot}_{\mu\nu}=g_{\mu\nu}+h_{\mu\nu}\,.$
The method that we will use for the stability analysis is the same as the one developed by Regge and Wheeler back in 1957 \cite{Regge:1957td} and which later corrected and enriched by Zerilli and Vishveshwara in 1970 \cite{Zerilli:1970se,Vishveshwara:1970cc,Zerilli:1971wd}. However, since we are dealing with a more general Lagrangian density we also need to incorporate the perturbations of the scalar field $\Phi$, namely
\eq$\label{Phi-tot}
\Phi^{tot}=\Phi+\del\Phi\,.$
The perturbations are distinguished into two distinct classes based on their parity: perturbations with \textit{odd} parity $(-1)^{L+1}$  also known as axial, and perturbations with \textit{even} parity $(-1)^L$ also known as polar, where $L$ is the angular momentum of the particular perturbation mode. Although our black-hole solution is static, the perturbations depend on all spacetime coordinates. Using the method of separation of variables, the decomposition into modes with fixed energy is accomplished via the term $\exp(-ikt)$, with $k$ being the frequency of the mode, while the decomposition into modes with fixed angular momentum $L$ is achieved via the tensor spherical harmonics \cite{Mathews:1962,Zerilli-3,Zerilli:1971wd,Moncrief:1974am,Stewart:1974uz,Thorne:1980ru} which generalize the well-known spherical harmonics $Y_L^{M_L}(\theta,\varphi)$.
Our theory is invariant under diffeomorphisms  and therefore we may use the gauge freedom in order to simplify the components of the perturbation tensor $h_{\mu\nu}$. By using the Regge-Wheeler gauge \cite{Regge:1957td} and setting to zero the z-component of the angular momentum $M_L$,\,
we eliminate the dependence on the $\varphi$ coordinate. One is allowed to specialize the z-component of the angular momentum, namely $M_L=0$, since the physics of the perturbations will not be altered by this choice. Consequently, we are led to the Legendre polynomials $P_L(\cos\theta)$ which are equal to the spherical harmonics $Y_L^0(\theta,\varphi)$. In this gauge, the perturbations are called canonical and they have the following form: 
\begin{itemize}
    \item Odd perturbations
\eq$\label{P-odd}
h^{\rm odd}_{\mu \nu}= \left[
\begin{array}{cccc} 
0 & 0 &0 & h_0(r) 
\\ 0 & 0 &0 & h_1(r)
\\ 0 & 0 &0 & 0
\\ h_0(r) & h_1(r) &0 &0
\end{array}\right] e^{-i k t}
\sin\theta\,\partial_\theta P_L(\cos\theta)\,,\hspace{2em}\del\Phi=0\,,$
\item and even perturbations
\eq$\label{h-even}
h^{\rm even}_{\mu \nu}= \left[
 \begin{array}{cccc} 
 H_0(r) e^{A(r)}B(r) & H_1(r) &0 & 0 
\\ H_1(r) & H_2(r)/B(r)  &0 & 0
\\ 0 & 0 &r^2 K(r) & 0
\\ 0 & 0 &0 & r^2 K(r)\,\sin^2\theta
\end{array}\right] e^{-i k t}
P_L(\cos\theta)\,,\\[2mm]$
\eq$\label{phi-even}
\del\Phi(r)=\widetilde{\Phi}(r)\,e^{-i k t}P_L(\cos\theta)\,.$
\end{itemize}
In the above, we have used the same notation as in the Regge and Wheeler work \cite{Regge:1957td} while the expressions of $A(r)$ and $m(r)$ are given by Eqs. \eqref{A-r} and \eqref{m-r}.  

In this work, we will focus only on the odd perturbations. The even perturbations, will be studied separately in a future work, since, due to their complexity, we have not been able to obtain rigorously an argument which decides either for or against the stability of our solution. In Appendix \ref{app: even-pert} we present the four by four system of differential equations which characterizes the polar perturbations.

Substituting now Eqs. (\ref{metr-tot}-\ref{P-odd}) to the field equations and keeping only the linear terms in $h_{\mu\nu}$, we find that for the odd metric perturbations there are only two independent equations: the $(\theta,\varphi)$ equation
\begin{equation}
    \left\{\frac{e^A}{4r^2} \left[h_1(A'B+2B')+2B\,h_1')\right]-\frac{ik}{2r^2B}\, h_0 \right\} \left[2 \cot \theta\, \pa_\theta P_L(\cos\theta)+L  (L +1) P_L(\cos\theta)\right]=0\,,\label{perteq1}
\end{equation}
and the $(r,\varphi)$ equation
\begin{align}
    &\bigg\{ h_1 \bigg[\frac{e^A}{4r^4} \left(3 r^2 A' B'+2 r^2 B''+2 L  (L +1)\right)+\frac{e^A B}{2r^4} \left(r^2 A''+\frac{r^2 A'^2}{2}- r A'+ r^2 \Phi'^2-2\right)- \frac{k^2}{r^2B}\bigg]\nonumber\\[2mm]
   &\hspace{0.5em}- \frac{i k}{2r^2B}\, h'_0+\frac{ ik}{r^3B}\, h_0  \bigg\}\pa_\theta P_L(\cos\theta)=0.\label{perteq2}
\end{align}
Notice here that for $L=0$, $P_0(\cos\theta)=1$, and both Eqs. \eqref{perteq1}, \eqref{perteq2} become identically zero. Thus, for $L=0$ there are no odd-parity perturbations in the canonical ansatz. The same result also holds for $L=1$, for which $P_1(\cos\,\theta)=\cos\,\theta$ and Eq. \eqref{perteq1} is satisfied identically. In this case, although there is Eq. \eqref{perteq2} to work with, by performing the coordinate transformation
\eq$\label{new-gauge}
x'^\mu=x^\mu+i\, \del^{\mu}{}_\varphi\, \frac{ e^{-ikt}}{k\,r^2}\,h_0(r)\,,$
and redefining the function $h_1(r)$ through the relation
\eq$\label{h1-redef}
h_1(r)=i\,\frac{rh_0'(r)-2h_0(r)}{k\,r}\,,$
we can make Eq. \eqref{perteq2} identically zero as well. Thus, in the case of $L=1$ the perturbations can be gauged away completely. Therefore, the odd-parity modes exist only for $L\geq 2$. Since the angular part of the above equations is non-vanishing or non-singular for $L\geq 2$ we may focus only on their radial part. Eq. (\ref{perteq1}) may be easily solved for $h_0(r)$ and then by substituting the result into Eq. (\ref{perteq2}) we get:
\begin{equation}
    \label{h1-dif}
    h_1''(r)+q_1(r)h_1'(r)+\left[ k^2 q_k(r)+q_0(r)  \right]h_1(r)=0\,,
\end{equation}
where
\begin{align}
    q_1(r)&=\frac{3 A'}{2}+\frac{3 B'}{B}-\frac{2}{r}\,, \qquad\quad
    q_k(r)=\frac{ e^{-A}}{B^2}\,,\\[3mm]
    q_0(r)&=\frac{ B'^2}{B^2}-\frac{A''}{2}-\frac{r B' \left(4-r A'\right)+2 L  (L +1)}{ 2r^2B}+\frac{2}{r^2}-  \Phi'^{\,2}\,.\label{perteq3}
\end{align}
It is clear now that Eq. (\ref{h1-dif}) determines the dynamics of the system while $h_0$ is a dependent function. For $\Phi=0$ and $V(\Phi)=0$ the above equation reduce to the Regge-Wheeler equation \cite{Regge:1957td}. Here we are interested only in the stability of the system and therefore we do not have to solve Eq. (\ref{h1-dif}). Instead, due to the time evolution factor $\exp(-i k t)$, we have only to determine whether or not the frequency $k$ is purely imaginary. To do this we have to eliminate the term with the first derivative in the above equation and thus bring Eq. \eqref{h1-dif} into a Schrödinger-like form. To this end, we introduce a new perturbation function $\Psi(r)$ through the relation 
\begin{equation}
    h_1(r)=\dfrac{r\, \Psi(r)}{B(r)\, e^{A(r)/2}}\,,
\end{equation}
and we also impose the tortoise coordinate $r^*$ via the transformation $dr^*=dr\,e^{-A(r)/2}/B(r)$. The tortoise coordinate transforms the region $[r_h,+\infty)$ to $(-\infty,+\infty)$ and therefore it parametrizes the whole exterior spacetime of the black hole. Also we observe that the coefficients $q_k(r)$ and $q_0(r)$  diverge at the horizon. The introduction of both the new function $\Psi$ and the tortoise coordinate $r^*$ eliminate this divergences and transform Eq. (\ref{h1-dif}) to the Schrödinger-like form: 
\begin{equation}
\label{Schr-eq}
    \frac{d^2\Psi(r^*)}{dr^{*2}}+\left[ k^2 - \mathcal{V}(r)   \right]\,\Psi(r^*)=0\,,
\end{equation}
\begin{figure}[t]
    \centering
    \begin{subfigure}[b]{0.49\textwidth}
    \includegraphics[width=1\textwidth]{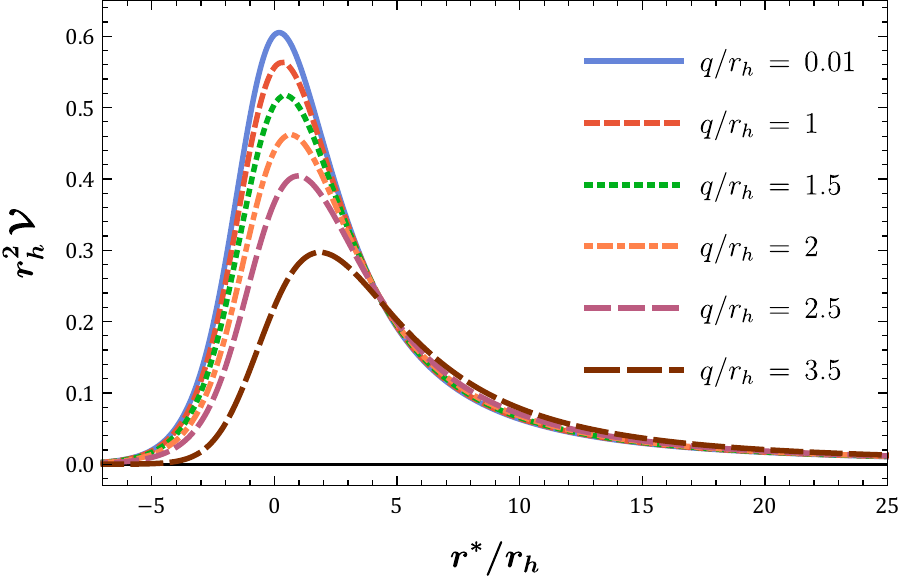}
    \caption{\hspace*{-2.8em}}
    \label{subf: Stab1}
    \end{subfigure}
    \hfill
    \begin{subfigure}[b]{0.49\textwidth}
    \includegraphics[width=1\textwidth]{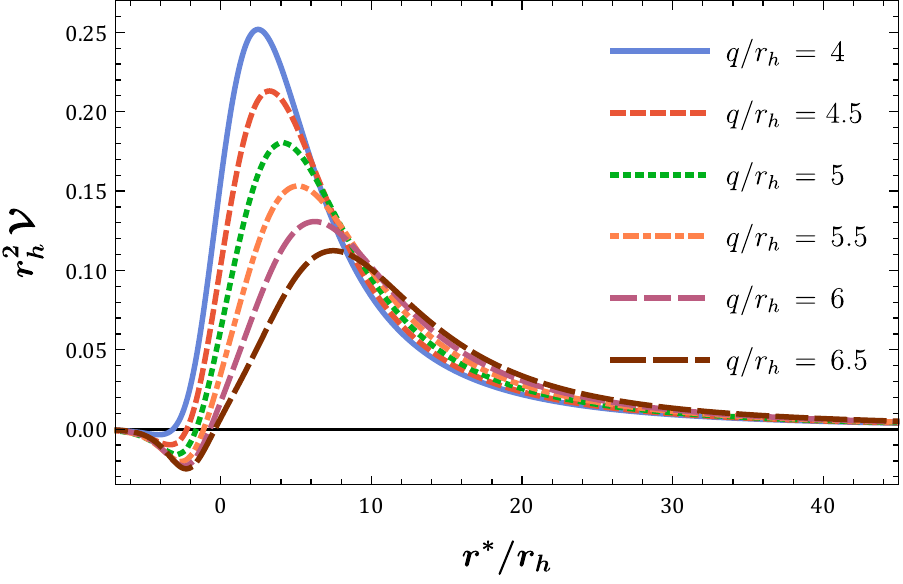}
    \caption{\hspace*{-3.7em}}
    \label{subf: Stab2}
    \end{subfigure}
    \caption{The potential $r_h^2\,\mathcal{V}$ in terms of the radial coordinate $r^*/r_h$ for (a) six stable and (b) six unstable solutions for   
      different values of the dimensionless parameter $q/r_h$, and angular momentum $L=2$.}
    \label{fig: stab}
\end{figure}
where the potential $\mathcal V(r)$ is given by 
\begin{equation}
    \mathcal{V}(r)=\frac{e^A B}{2r} \left\{B' \left(3 r A'-2\right)+B \left[2 r \left(A''+\Phi'^{\,2}\right)+r
   A'^2-3 A'\right]+2 r B''+\frac{2 L  (L +1)}{r}\right\}\,.\label{potsch}
\end{equation}
In the tortoise coordinate both the new dynamical function $\Psi$ and the potential $\mathcal{V}$ are everywhere regular outside the horizon. Also, it may be shown that the potential tends to zero in both asymptotic regions.

\begin{figure}[t]
    \centering
    \includegraphics[width=0.5\textwidth]{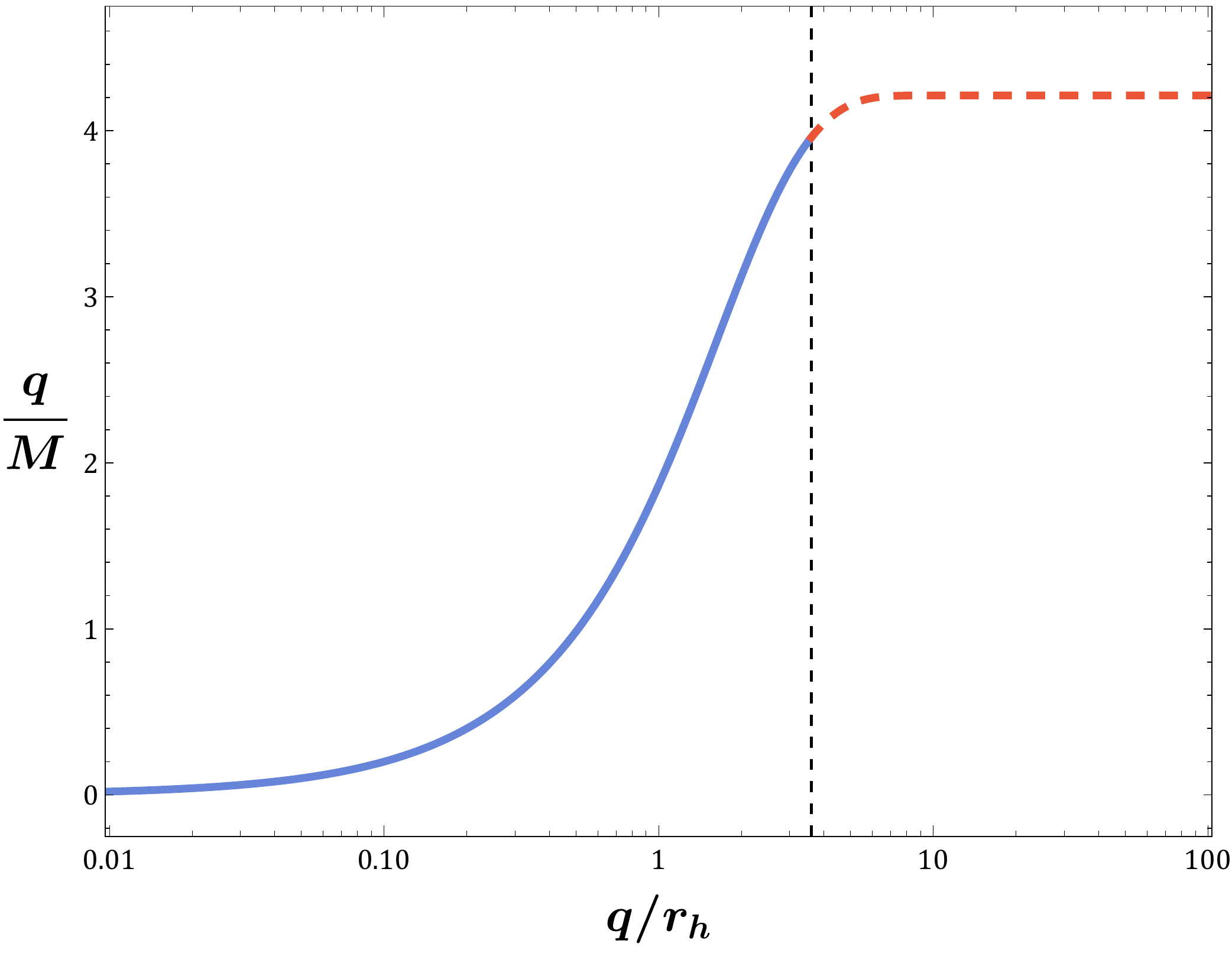}
    \caption{The graph of the ratio $q/M$ in terms of $q/r_h$. Both quantities are dimensionless, while the horizontal axis is logarithmic and the vertical dashed line lies at $q/r_h=3.6$. The values of $q/M$ correspond to black-hole solutions. The blue continuous line represents stable solutions, while the red dashed line refers to unstable ones.}
    \label{fig: q/M}
\end{figure}

An unstable mode corresponds to a bound state of the Schrödinger equation \eqref{Schr-eq} i.e. to a negative eigenvalue $k^2<0$. This means that the frequency $k$ is purely imaginary and therefore the mode grows exponentially due to the term $\exp(-ikt)$. For a potential $\mathcal{V}(r^*)$ that vanishes in $r^*\rightarrow\pm\infty$, the condition for at least a bound state is \cite{Buell}
\begin{equation}
    \int_{-\infty}^{+\infty}\mathcal{V}(r^*)\,dr^*<0\,.
\end{equation}
However, as the authors in \cite{Buell} argue, even if the above integral is positive a bound state could still exist. Intuitively this makes sense, since for a potential with a shape as the ones depicted in Fig. \myref{fig: stab}{subf: Stab2}, there is nothing which prevents a bound state to exist in the region where the potential is negative-definite and has the shape of a well. Consequently, we claim that if the potential $\mathcal{V}$ is negative in a region, then this particular solution contains at least one unstable mode. The term $L(L+1)$ in Eq. (\ref{potsch}) adds a positive angular barrier in the potential. Therefore it is sufficient to examine the stability of our black hole solutions on the mode with the smallest possible value for the angular momentum, i.e. $L=2$.  In Fig. \ref{fig: stab} we depict the potential $r_h^2\,\mathcal{V}(r^*)$ for twelve black hole solutions for different values of the parameter $q/r_h$. We find that the solutions with $q/r_h<3.6$  are stable while the solutions with $q/r_h\geq 3.6$ are unstable. Using Eq. \eqref{rh/M}, we can evaluate the ratio $q/M$, regarding black-hole solutions, for any given value of the dimensionless quantity $q/r_h$. The graph of the ratio $q/M$ in terms of the quantity $q/r_h$ is given in Fig. \ref{fig: q/M}. The vertical line which lies at $q/r_h=3.6$ distinguishes the stable from the unstable black-hole solutions, while for $q/r_h=3.59$ the quantity $q/M$ takes its highest value, which is approximately 3.956. Substituting $q/r_h=3.59$ to Eqs. (\ref{rh/M}) and (\ref{omega-R}) we find that the most compact stable black hole is 0.551 times smaller than the Schwarzschild one, while it rotates 2.491 times faster compared to the slow-rotating Schwarzschild black hole.

\section{Epilogue}\label{sec4}

In this work, we have considered a very simple theory which contains a scalar field with its kinetic and potential terms minimally coupled to the gravitational field. We have assumed a spherically symmetric form for the metric tensor and a coulombic form for the scalar field. Consequently, performing the variation of the action functional with respect to the fields of our theory, we have obtained the field equations which then have been explicitly solved to determine the analytic expressions of the spacetime line-element and the self-interacting scalar potential. The spherical symmetric solution which was considered in this work had been firstly appeared in \cite{Herdeiro:2015waa}.

We started our analysis by examining the properties of the spacetime geometry. By taking the expansion of the metric components at infinity, we were able to show that the derived spacetime geometry is asymptotically flat, while---as the scalar curvature quantities dictate---a true spacetime singularity is present at $r=0$. Studying the causal structure of the background geometry we have also shown that the spacetime admits one horizon, hence, the derived solution describes the geometry of a black hole. However, by letting the scalar hair $q$---which is of secondary type---to obtain arbitrarily large values compared to the black-hole mass $M$, we encounter a transition from a black hole to a naked singularity. The critical value between the scalar hair and the black-hole mass in which this transition takes place is $q/M\gtrsim 4.2$. For any combination of the parameters $q$ and $M$ for which the black-hole horizon is apparent, the horizon radius of our solution is found to be always smaller than the horizon radius of the corresponding Schwarzschild black hole with the same mass. An appropriate choice of parameters $q$ and $M$ can lead to extremely low values of the ratio $r_h/(2M)$, and thus to ultra-compact black-hole solutions. However, examining the ultra-compact solutions from a thermodynamical point of view, we can readily deduce that as the horizon radius decreases the horizon entropy also decreases and therefore the ultra-compact black holes are thermodynamically less stable. 

The most interesting property of electrovacuum black-hole solutions in General Relativity is their simplicity and uniqueness. It is well-known that in the context of GR black hole solutions are determined only by three physical quantities: \textit{mass, electromagnetic charge and angular momentum}. The fact that the association of GR black holes with any other conserved ``charge" or quantum number is forbidden, is a direct result of the no-hair theorems. Even in the context of scalar-tensor theories of gravity, no-scalar hair theorems have been also formulated. In this case though they prohibit the association of black-hole solutions with scalar hair. As we already mentioned, the term scalar hair has the sense of a non-trivial scalar configuration that accompanies the black hole, and not just a conserved scalar charge.  However, it was shown that although the no-hair theorems have a catholic validity, the no-scalar hair theorem is only viable in a subclass of the scalar-tensor theories. The  theory which we considered in this work constitute a simple but a special type of model, which for negative-definite scalar potentials can evade the no-scalar hair theorem and lead to hairy and most importantly regular black-hole solutions. As we illustrated in Section \ref{sec1} the scalar potential in our case exhibits the desired behaviour, $V(\Phi)<0$, which is necessary for the evasion of the no-scalar hair theorem, hence, our hairy black hole solutions are completely justified. Having in our disposal the explicit expressions of the scalar field and the potential of the field, the evaluation of the energy-momentum tensor associated with the scalar field is an easy task. Then, from the mixed components of the stress-energy tensor it is straightforward to identify the energy density $\rho_E$ and the pressures $p_r$ (radial pressure) and $p_\theta$ (tangential pressure). Having their analytic expressions, we have shown that the strong energy conditions are satisfied in the region which lies outside the black hole horizon, namely the causal region of the spacetime, while beyond the black-hole horizon, in the interior of the black hole, the energy conditions are violated. This is indeed anticipated since the assumed scalar field was a coulombic one which by its turn led to a negative-definite scalar potential that decreases as we move closer to the singularity.

Apart from the description of static solutions, we were able to construct solutions which are slowly rotating. To achieve this, we have treated the rotation of the black hole as axisymmetric perturbations on the static and spherically symmetric background metric. By doing so, we obtained new differential equations for the angular velocity of the black hole, which together with the appropriate boundary condition have been solved analytically. As it is anticipated, the resulting angular velocity was found to depend only on the radial component $r$, while its magnitude compared to the angular velocity of a slow-rotating and equally massive Schwarzschild black hole was found to be always larger. Intuitively, this result should not be surprising, since, as we have discussed previously, our solution describes ultra-compact black holes with horizon radii always smaller than the horizon radius of the corresponding Schwarzschild black hole with the same mass. Hence, assuming that both solutions could emerge from the collapse of the same star, the one with the smaller horizon radius is expected to spin faster. As it has been illustrated in Section \ref{sec2}, for values $q/r_h\geq 7$, the ratio of the angular velocity of the slow-rotating ultra-compact black hole to the angular velocity of the slow-rotating Schwarzschild black hole remains constant, and specifically $\omega(r_h)/\omega_{\rm Sch}(2M)\simeq 3.2$. However, so fast rotation lies beyond the slope of the slow rotation approximation and therefore one should take into account the next to leading order terms, or perform a non-perturbative rotation analysis.

Modified theories of gravity have been extensively studied over the past years, and as a result, the literature of hairy black holes has been rapidly enhanced with novel solutions. However, it is not clear if all of them can be considered as astrophysical objects. The way to decide if a black-hole solution on the paper can be seriously taken into account as an astrophysical object is via its stability under spacetime perturbations. This is the reason why stability analysis is such a requisite and crucial part of any complete study of the physical characteristics of a black-hole solution. As far as our solution is concerned, we have studied its stability under axial perturbations, and we have derived explicitly the Schr\"{o}dinger-like equation and the effective potential. It is solely the behaviour of the effective potential which decides whether the solution is stable or not. Plotting the effective potential in terms of the dimensionless coordinate $r^*/r_h$---with $r^*$ being the tortoise coordinate---for various values of the dimensionless parameter $q/r_h$, we have found that for $q/r_h\leq 3.59$, all resulting black-hole solutions are stable under axial perturbations. For the critical case of $q/r_h=3.59$, one can easily determine that $r_h/(2M)=0.551$, $q/M=3.956$, and $\omega(r_h)/\omega_{\rm Sch}(2M)=2.491$. Thus, according to our analysis \textit{the most compact and stable black hole solution is 0.551 times smaller than the Schwarzschild one, while it rotates 2.491 times faster compared to the slow-rotating Schwarzschild black hole.}

The axial perturbations may provide us an indication about the stability of our solutions. However, the full stability of the system may be determined by the examination of both axial and polar perturbations. Therefore, future directions of our work could be the examination of the polar perturbations, which were left aside in the context of this work due to their complexity. In order to fully analyze the even perturbations, one needs advanced mathematical methods since a system of four first order differential equations with non-constant coefficients is not an easy task to undertake. Having attained the complete stability analysis of a particular black-hole solution, one can then study its quasi-normal modes (QNMs), which are directly related to the way a black hole oscillates. Since different solutions is most likely to have different frequency spectra, knowing the QNMs of a compact object is like knowing the digits of a person’s ID card. Therefore, it is very likely in the future to discover new compact objects, if they exist, through their quasi-normal modes from gravitational wave experiments \cite{Berti:2018vdi}. Note here that the future updates of the LIGO-Virgo experiments are expected to probe these frequencies, hence, the study of QNMs is of great importance. Apart from the detailed study of spacetime perturbations and QNMs, a thorough investigation of the asymptotically (A)dS solutions---which we only discussed briefly in Appendix \ref{app: ads}---could also be a very interesting path for one to take. These types of solutions possess an effective cosmological constant and therefore could also be used as models for dark energy. It is in our future plans to return to, at least, one of those questions.

{\bf Acknowledgements.} We would like to thank Panagiota Kanti for useful comments at the final stages of this work. A.B. is also thankful to Christos Charmousis for many enlightening discussions. The research of T.N. was co-financed by Greece and the European Union (European Social Fund- ESF) through the Operational Programme “Human Resources Development, Education and Lifelong Learning” in the context of the project “Strengthening Human Resources Research Potential via Doctorate Research – 2nd Cycle” (MIS-5000432), implemented by the State Scholarships Foundation (IKY). The authors happily acknowledge networking support by the GWverse COST Action CA16104, “Black holes, gravitational waves and fundamental physics".

\appendix

\vspace*{1em}

\newpage

\section{Asymptotically (Anti-)de Sitter black hole solution}
\label{app: ads}

The most general solution of the system of differential equations \eqref{eq1}-\eqref{eq3} is 

\begin{equation}
    A(r)=-\frac{q^2}{4r^2},
\end{equation}
\begin{align}
   m(r)= \frac{r}{2}+\frac{4 r^3}{q^2} &+ \frac{r^2 e^{\frac{q^2}{8 r^2}}}{q^2}\left[-12 M+   \sqrt{2 \pi } \,q \,\erf\left(\frac{q}{2 \sqrt{2}\, r}\right) \right]\nonumber\\[2mm]
   &-\frac{r^3 e^{\frac{q^2}{4 r^2}}}{q^3} \left\{4 q -\frac{q^3\Lambda_{\rm eff}}{6} - 12 \sqrt{2 \pi } \,M \,\erf\left(\frac{q}{2 \sqrt{2}\, r}\right)+\pi q
   \left[\erf\left(\frac{q}{2 \sqrt{2}\, r}\right)\right]^2  \right\},
\end{align}
\begin{align}
    V(\Phi)=&\frac{2(\Phi ^2+24)}{q^2}-\frac{12 e^{\frac{\Phi ^2}{8}} \Phi }{q^3}\left[ 12 M-\sqrt{2 \pi }\, q\, \text{erf}\left(\frac{\Phi }{2 \sqrt{2}}\right)   \right]\nonumber\\[2mm]
    &+\frac{e^{\frac{\Phi ^2}{4}} \left(\Phi ^2-12\right)}{q^3}\left\{4 q - \frac{q^3  \Lambda_{\rm eff}}{6} - 12 \sqrt{2 \pi }\, M \,\erf\left(\frac{\Phi }{2 \sqrt{2}}\right) +  \pi  q
   \left[\erf\left(\frac{\Phi }{2 \sqrt{2}}\right)\right]^2   \right\},
\end{align}
where 
\begin{equation}
    \Lambda_{\rm eff}=\frac{24}{q^2}+6\,C.
\end{equation}
The constant $C$ in the definition of the $\Lambda_{\rm eff}$ is an integration constant from Eq. (\ref{eq2}). By taking the expansion at infinity of the metric function $B(r)=1-2m(r)/r$ we find that the spacetime assumes an (A)dS form
\begin{equation}
  B(r)=  -\frac{\Lambda_{\rm eff} }{3}r^2+\left(1-\frac{q^2\Lambda_{\rm eff} }{12}\right)-\frac{2 M}{r}+\frac{q^2}{4r^2}\left(1-\frac{q^2\Lambda_{\rm eff}}{24}\right)+\mathcal{O}\left( \frac{1}{r^3}\right).
\end{equation}
Note that at infinity, where the scalar field $\Phi=q/r$ vanishes, the potential has the following expansion
\begin{equation}
    V(\Phi)=2 \Lambda_{\rm eff}+\frac{\Lambda_{\rm eff}\,\Phi^2}{3}+\frac{\Lambda_{\rm eff}\,\Phi^4}{48}-\frac{2M\,\Phi^5}{5q^3}+\mathcal{O}(\Phi^6)
\end{equation}
Finally, if we ignore the effective cosmological constant $\Lambda_{\rm eff}=0$ we get the asymptotically flat solution given in Eqs. (\ref{A-r}-\ref{V-phi}).

\section{Analytic expressions of scalar curvature quantities}
\label{app: sc-curv}

Below, we present the analytic expressions of the curvature invariant quantities $R\equiv g^{\mu\nu}R_{\mu\nu}$, $\mathcal{R}\equiv R^{\mu\nu}R_{\mu\nu}$ and $\mathcal{K}\equiv R^{\mu\nu\kappa\lam}R_{\mu\nu\kappa\lam}$ which result from the line-element \eqref{metr-ans}.
\bal$\label{Ricci}
R=\frac{96}{q^2}&+\frac{3\, e^{\frac{q^2}{4 r^2}}}{q^3\,r^2} \left(q^2-8 r^2\right) \left\{4 q-12 \sqrt{2 \pi }\, M\, \text{erf}\left(\frac{q}{2 \sqrt{2}\, r}\right)+\pi  q\, \text{erf}\left[\left(\frac{q}{2 \sqrt{2}\, r}\right)\right]^2\right\}\nonum\\[1mm]
&-\frac{e^{\frac{q^2}{8 r^2}}}{q^2\, r^3} \left(q^2-24 r^2\right) \left[\sqrt{2 \pi }\, q\,\text{erf}\left(\frac{q}{2 \sqrt{2} r}\right)-12 M\right]\,,$
\bal$\label{sq-Ricci}
\mathcal{R}&=\frac{12}{q^4r^6}\Bigg\{r^2 \left(q^4+192 r^4\right)-6 M\, r\, e^{\frac{q^2}{8 r^2}} \left(q^4-8 q^2 r^2+192 r^4\right)+4 r^2 e^{\frac{q^2}{2 r^2}} \left(q^4-12 q^2 r^2+48 r^4\right)\nonum\\[1mm]
&\hspace{1em}+12 M\, r\, e^{\frac{3 q^2}{8 r^2}} \left(q^4-16 q^2 r^2+96 r^4\right)+12 e^{\frac{q^2}{4 r^2}} \left[ M^2 q^4+4 r^4 \left(36 M^2+q^2\right)\right]\nonum\\[1mm]
&\hspace{1em}-2r^2 e^{\frac{q^2}{4 r^2}} \left[q^2  \left(72 M^2+q^2\right)+192 r^4\right]\Bigg\}+\frac{\sqrt{\pi }\, e^{\frac{q^2}{8 r^2}}}{q^6r^6} \text{erf}\left(\frac{q}{2 \sqrt{2} r}\right)\Bigg\{6 \sqrt{2}\, q^3 r \left(q^4-8 q^2 r^2+192 r^4\right)\nonum\\[1mm]
&\hspace{1.2em}+2 q\, e^{\frac{q^2}{8 r^2}} \left(q^6-15 q^4 r^2+216 q^2 r^4-576 r^6\right)\left[\sqrt{\pi }\, q\, \text{erf}\left(\frac{q}{2 \sqrt{2} r}\right)-12 \sqrt{2} M\right]\nonum\\[1mm]
&\hspace{1.2em}-3 q r e^{\frac{q^2}{4 r^2}} \left(q^4-16 q^2 r^2+96 r^4\right) \Bigg[\sqrt{\pi }\, q\, \text{erf}\left(\frac{q}{2 \sqrt{2} r}\right) \left(\sqrt{2 \pi }\, q\, \text{erf}\left(\frac{q}{2 \sqrt{2} r}\right)-36 M\right)\nonum\\[1mm]
&\hspace{1.2em}+4 \sqrt{2} \left(36 M^2+q^2\right)\Bigg]+3 r^2 e^{\frac{3 q^2}{8 r^2}} \left(q^4-12 q^2 r^2+48 r^4\right) \Bigg[\sqrt{\pi }\,\text{erf}\left(\frac{q}{2 \sqrt{2} r}\right) \Bigg(288 M^2\nonum\\[1mm]
&\hspace{1.2em}+q^2 \left(\pi \bigg[\text{erf}\left(\frac{q}{2 \sqrt{2} r}\right)\bigg]^2+8\right)-24 \sqrt{2 \pi }\, M q\, \text{erf}\left(\frac{q}{2 \sqrt{2} r}\right)\Bigg)-96 \sqrt{2}\, M q\Bigg]\Bigg\}\,,$
\bal$\label{Riem-sc}
\mathcal{K}&=\frac{24}{q^4r^4}\Bigg\{r^2\left(q^4+64r^4\right)-2 M\, r\, e^{\frac{q^2}{8 r^2}} \left(3 q^4-8 q^2 r^2+192 r^4\right)+2 r^2 e^{\frac{q^2}{2 r^2}} \left(q^4-8 q^2 r^2+32 r^4\right)\nonum\\[1mm]
&\hspace{1.2em}+8 M\, r\, e^{\frac{3 q^2}{8 r^2}} \left(q^4-8 q^2 r^2+48 r^4\right)-2 e^{\frac{q^2}{4 r^2}} \left[r^2 \left(q^4-8 q^2 r^2+64 r^4\right)-6 M^2 \left(q^4-4   q^2 r^2+48 r^4\right)\right]\Bigg\}\nonum\\[1mm]
&\hspace{1.2em}+\frac{\sqrt{\pi }\, e^{\frac{q^2}{8 r^2}}}{q^6r^6} \text{erf}\left(\frac{q}{2 \sqrt{2} r}\right)\Bigg\{4 \sqrt{2}\, q^3 r \left(3 q^4-8 q^2 r^2+192 r^4\right)\nonum\\[1mm]
&\hspace{1.2em}+4 q\, e^{\frac{q^2}{8 r^2}} \left(q^6-7 q^4 r^2+72 q^2 r^4-192 r^6\right) \left[\sqrt{\pi} q\, \text{erf}\left(\frac{q}{2 \sqrt{2} r}\right)-12 \sqrt{2} M\right]\nonum\\[1mm]
&\hspace{1.2em}-4 q\, r\, e^{\frac{q^2}{4 r^2}} \left(q^4-8 q^2 r^2+48 r^4\right) \Bigg[\sqrt{\pi }\, q\, \text{erf}\left(\frac{q}{2 \sqrt{2} r}\right) \left(\sqrt{2 \pi }\, q \,\text{erf}\left(\frac{q}{2 \sqrt{2} r}\right)-36 M\right)\nonum\\[1mm]
&\hspace{1.2em}+4 \sqrt{2} \left(36 M^2+q^2\right)\Bigg]+3 r^2 e^{\frac{3 q^2}{8 r^2}} \left(q^4-8 q^2 r^2+32 r^4\right) \Bigg[\sqrt{\pi }\, \text{erf}\left(\frac{q}{2 \sqrt{2} r}\right) \Bigg(288 M^2\nonum\\[1mm]
&\hspace{1.2em}+q^2 \left(\pi \bigg[\text{erf}\left(\frac{q}{2 \sqrt{2} r}\right)\bigg]^2+8\right)-24 \sqrt{2 \pi }\, M\, q\, \text{erf}\left(\frac{q}{2 \sqrt{2} r}\right)\Bigg)-96\sqrt{2}\, M\, q\Bigg]\Bigg\}\,.$

\newpage
\section{Analytic expressions of the mixed stress-energy tensor components}
\label{app: ene-mom}

From eq. \eqref{stress-ten} it is straightforward to deduce that the independent mixed stress-energy tensor components are the energy density $\rho_E=-T^{(\Phi)t}{}_t$, and the radial pressure $p_r=T^{(\Phi)r}{}_r$. Their analytic expressions can be calculated from \eqref{stress-ten} with the use of eqs. \eqref{metr-ans}, \eqref{A-r}-\eqref{V-phi}, and they are presented below.
\bal$\label{en-dens}
\rho_E(r)=&\,\frac{1}{q^2r^3}\left[6 M \,e^{\frac{q^2}{8 r^2}} \left(q^2-12 r^2\right)+4 r\, e^{\frac{q^2}{4 r^2}} \left(q^2-6 r^2\right)-q^2 r+24 r^3\right]\nonum\\[2mm]
&-\frac{\sqrt{2 \pi }}{2q^3r^3}\,e^{\frac{q^2}{8 r^2}} \,\text{erf}\left(\frac{q}{2 \sqrt{2} r}\right) \left[-12 q^2 r \left(r-2 M e^{\frac{q^2}{8 r^2}}\right)-144 M r^3 e^{\frac{q^2}{8 r^2}}+q^4\right]\nonum\\[2mm]
&+\frac{\pi\left(q^2-6 r^2\right)}{q^2r^2}\,   e^{\frac{q^2}{4 r^2}}\,  \left[\text{erf}\left(\frac{q}{2 \sqrt{2} r}\right)\right]^2\,,\\[7mm]
\label{rad-press}
p_r(r)=&\, \frac{3}{q^2r^3}\left[2 M\, e^{\frac{q^2}{8 r^2}} \left(q^2+12 r^2\right)-r \left(q^2+8 r^2\right)+8 r^3 e^{\frac{q^2}{4 r^2}}\right]\nonum\\[2mm]
&-\frac{\sqrt{2 \pi }}{2q^3r^3}\, e^{\frac{q^2}{8 r^2}}\, \text{erf}\left(\frac{q}{2 \sqrt{2} r}\right) \left(144 M r^3 e^{\frac{q^2}{8 r^2}}+q^4+12 q^2 r^2\right)+\frac{6 \pi }{q^2}\, e^{\frac{q^2}{4 r^2}}\, \left[\text{erf}\left(\frac{q}{2 \sqrt{2} r}\right)\right]^2\,.$
Combining the above relations one may compute that
\bal$
\rho_E(r)+p_r(r)=&\,\frac{4}{r^3}\left(3 M\, e^{\frac{q^2}{8 r^2}}+r\, e^{\frac{q^2}{4 r^2}}-r\right)-\frac{\sqrt{2 \pi }}{qr^3}\, e^{\frac{q^2}{8 r^2}}\, \text{erf}\left(\frac{q}{2 \sqrt{2} r}\right) \left(12 M r\, e^{\frac{q^2}{8 r^2}}+q^2\right)\nonum\\[2mm]
&+\frac{\pi  }{r^2}\, e^{\frac{q^2}{4 r^2}} \left[\text{erf}\left(\frac{q}{2 \sqrt{2} r}\right)\right]^2\,.$
Using now eq. \eqref{rh/M} in the above relation, we find that
\bal$
\rho_E(r)+p_r(r)=&\,\frac{1}{r^2}\,e^{\frac{q^2}{4 r^2}}\left\{4+\pi\left[\text{erf}\left(\frac{q}{2 \sqrt{2} r}\right)\right]^2\right\}-\frac{1}{r^3}\left[4r+\sqrt{2\pi}\,q\,e^{\frac{q^2}{8 r^2}}\,\text{erf}\left(\frac{q}{2 \sqrt{2} r}\right) \right]\nonum\\[2mm]
&-\frac{r_h\,e^{\frac{q^2}{4 r_h^2}}\left\{4+\pi\left[\text{erf}\left(\frac{q}{2 \sqrt{2} r_h}\right)\right]^2\right\}e^{\frac{1}{8} q^2
   \left(\frac{1}{r^2}-\frac{1}{r_h^2}\right)}}{r^3 \left[q-\sqrt{2 \pi }\, r_h\, e^{\frac{q^2}{8 r_h^2}} \text{erf}\left(\frac{q}{2 \sqrt{2}
   r_h}\right)\right]}\left[q-\sqrt{2\pi}\,r\,e^{\frac{q^2}{8r^2}}\,\text{erf}\left(\frac{q}{\sqrt{2\pi}\,r}\right)\right]\nonum\\[2mm]
&+\frac{\left[4 r_h+\sqrt{2 \pi }\, q\, e^{\frac{q^2}{8 r_h^2}} \text{erf}\left(\frac{q}{2 \sqrt{2} r_h}\right)\right]e^{\frac{1}{8} q^2
   \left(\frac{1}{r^2}-\frac{1}{r_h^2}\right)}}{r^3 \left[q-\sqrt{2 \pi }\, r_h\, e^{\frac{q^2}{8 r_h^2}} \text{erf}\left(\frac{q}{2 \sqrt{2}
   r_h}\right)\right]}\left[q-\sqrt{2\pi}\,r\,e^{\frac{q^2}{8r^2}}\,\text{erf}\left(\frac{q}{\sqrt{2\pi}\,r}\right)\right]\,.$
From the above equation one can easily observe that $\rho_E(r_h)+p_r(r_h)=0$. Due to the fact that the root at $r_h$ is not of even order leads us to conclude that the quantity $\rho_E(r)+p_r(r)$ changes sign at the black-hole horizon. This may also be observed in Fig. \myref{fig: SEC}{subf: SEC1}.

\section{The differential equations of polar perturbations}
\label{app: even-pert}

Substituting Eqs. (\ref{metr-tot}, \ref{Phi-tot}, \ref{h-even}) to the field equations and keeping only the linear terms in $h_{\mu\nu}$ and $\del\Phi$, we find that for the even perturbations there are four independent equations: $(t,r)$, $(t,\theta)$, $(r,r)$ and $(r,\theta)$, while $(\theta,\varphi)$ results to $H_2(r)=H_0(r)$. The system of differential equations regarding the functions $H_0(r)$, $H_1(r)$, $K(r)$ and $\widetilde{\Phi}(r)$ is the following
\eq$\label{even-syst}
\left[ \begin{array}{c}
    H_0'(r)  \\ H_1'(r) \\ K'(r) \\ \widetilde{\Phi}'(r)
\end{array}\right]=
\boldsymbol{\mathcal{M}(r)}\left[ \begin{array}{c}
    H_0(r)  \\ H_1(r) \\ K(r) \\ \widetilde{\Phi}(r)
\end{array}\right]\,, \hspace{1.5em} \boldsymbol{\mathcal{M}(r)}\equiv\left[ \begin{array}{cccc}
    M_{11}(r) & M_{12}(r) & M_{13}(r) & M_{14}(r) \\
    M_{21}(r) & M_{22}(r) & M_{23}(r) & M_{24}(r) \\
    M_{31}(r) & M_{32}(r) & M_{33}(r) & M_{34}(r) \\
    M_{41}(r) & M_{42}(r) & M_{43}(r) & M_{44}(r) \\
\end{array}\right]\,,$
where
\bal$
&M_{11}(r)=\frac{1}{r}-A'(r)-\frac{B'(r)}{B(r)}\,, \hspace{1em}M_{12}(r)=\frac{i L(L+1)}{2 k r^2}-\frac{i k e^{-A(r)}}{B(r)}\,, \hspace{1em}M_{13}(r)=\frac{A'(r)}{2}+\frac{B'(r)}{2 B(r)}-\frac{1}{r}\,,\nonum\\[2mm]
&M_{14}(r)=-\frac{q}{2r^2}\,, \hspace{1em} M_{21}(r)=M_{23}(r)=-\frac{ik}{B(r)}\,, \hspace{1em} M_{22}(r)=-\frac{A'(r)}{2} -\frac{B'(r)}{B(r)}\,, \hspace{1em} M_{24}(r)=0\,,\nonum\\[2mm]
&M_{31}(r)=\frac{1}{r}\,,\hspace{1em} M_{32}(r)=\frac{iL(L+1)}{2kr^2}\,,\hspace{1em} M_{33}(r)=\frac{A'(r)}{2}+\frac{B'(r)}{2 B(r)}-\frac{1}{r}\,,\hspace{1em} M_{34}(r)=\frac{q}{2r^2}\,,\nonum\\[2mm]
&M_{41}(r)=-\frac{r A'(r)}{q}-\frac{r B'(r)}{q B(r)}-\frac{L(L+1)}{q B(r)}-\frac{q}{2 r^2}+\frac{2}{q}\,,\nonum\\[2mm]
&M_{42}(r)=-\frac{i L(L+1) A'(r)}{2 k q}+\frac{2 i k r e^{-A(r)}}{q B(r)}-\frac{i L(L+1) B'(r)}{2 k q B(r)}\,,\nonum\\[2mm]
&M_{43}(r)=-\frac{r^2 [A'(r)]^2}{2 q}+\frac{r A'(r)}{q}-\frac{2 k^2 r^2 e^{-A(r)}}{q [B(r)]^2}-\frac{r^2 [B'(r)]^2}{2 q [B(r)]^2}+\frac{r B'(r)+L(L+1)-2-r^2 A'(r) B'(r)}{q B(r)}\,,\nonum\\[2mm]
&M_{44}(r)=-\frac{A'(r)}{2}-\frac{B'(r)}{2 B(r)}-\frac{  r^2 \pa_\Phi V\left(\Phi\right)}{q B(r)}-\frac{2}{r}\,.\nonum$

\addcontentsline{toc}{section}{References}
\bibliography{Bibliography}{}
\bibliographystyle{utphys}

\end{document}